\documentclass[11pt,a4paper]{article}
\usepackage{jheppub_kim}
\usepackage{rotating}
\usepackage{graphicx,epsfig}
\usepackage{amsmath}
\usepackage {amssymb}
\usepackage{subfigure}
\usepackage{multirow}
\usepackage{filecontents,catchfile}
\usepackage{float}
\usepackage{perpage}
\usepackage{mathtools}
\MakeSorted{figure}
\MakeSorted{table}
\usepackage[utf8x]{inputenc}
\usepackage{relsize}
\usepackage{cleveref}
\usepackage{pgfplots}
\usepackage{array,multirow}
\usepackage{soul}
\usepackage{subfigure}
\usepackage{dsfont}
\usepackage{hyperref}
\usepackage{txfonts}
\usepackage{amsfonts}
\usepackage{newlfont}
\usepackage{times}
\usepackage{amssymb}
\usepackage{secdot}
\usepackage{cmap}
\usepackage[explicit]{titlesec }

\graphicspath{ {./images/} }
\titlespacing{\paragraph}{0pt}{0pt}{.5em}[]

\usepackage[font=scriptsize]{caption}

\title{\boldmath  The Magnetic Field Effect on Thermodynamics of Hot QCD Matter using Extensive and non-Extensive Statistics.}

\author[a]{Essam Tarek,}
\author[a]{M.M.Ahmed,}
\author[b]{Asmaa G. Shalaby}

\affiliation[a]{Physics Department, Faculty of Science, Helwan University, Helwan  Egypt.}
\affiliation[b]{ Physics Department, Faculty of Science, Benha university, Benha 13518, Egypt.}

\emailAdd{E.Tarek@science.helwan.edu.eg}
\emailAdd{asmaa.shalaby@fsc.bu.edu.eg}

\abstract{We study in detail the thermodynamics of the quantum chromodynamics (QCD) matter utilizing two different statistics, extensive and non-extensive. The thermodynamics such as (pressure, number density, energy density , entropy and magnetization) are determined from both statistics at zero and non-zero magnetic field, $eB = 0, 0.2, 0.3 \hspace{0.05cm}GeV^{2}$. The magnetic field effect appears by adding a vacuum contribution to the free energy along side the thermal contribution. The extensive thermodynamics can be emerged from the resonance hadron gas model which is incorporated in the present work.  Accordingly, we repeat our calculations at zero and non-zero magnetic field for the non-extensive statistics. The theoretical results of thermodynamical quantities calculated from both statistics are confronted to the lattice results which show a reasonable agreement with the extensive thermodynamics, but overrated in non-extensive thermodynamics at high temperature only and higher entropic index. The response of the QCD matter due to the magnetic field is additionally examined, and it is concluded that QCD matter is considered as a paramagnetic matter.}


\begin{document}
\maketitle
\flushbottom

\section{Introduction}
 One of the most interesting questions that, the scientists are trying to answer is "How the universe was created?". The answer might start with the theory of the big bang which states that, the universe was born due to a massive explosion creating fireball. The firball was composed of a dense and hot quarks and gluons plasma (QGP), then, the fireball began to expand and cool off and the elementary particles began to recombine. Due to freezing of QGP, the quarks and gluons combined creating the ordinary matter beginning with the main building block of the Hydrogen atom, protons, neutrons which after that they are categorized as hadrons. After more decreasing of the temperature, the nuclei created which in turn combined with each other creating the ordinary matter \cite{letessier2002hadrons}.\\
Many efforts have been done to study the phase transition between  the hadronic matter and QGP theoretically and experimentally. Starting with a collision of heavy ions or two hadrons accelerated to extremely high energies such as pp, and Pb-Pb collision at the Large Hadron Collider (LHC), and the Au-Au collision at Relativistic Heavy Ion Collider (RHIC) \cite{Bhalerao14}. Unfortunately, the QGP matter can exist for a very short period of time (time span) begins to expand leading to a recombination of quarks and gluons forming the hadronic matter again at a critical temperature $\mathrm{T_{c}}$ \cite{Csernai,Chaudhuri,Rafelski}.

 The freezing-out process passes through different stages, the thermal freeze-out and chemical freeze-out. 
 The chemical freeze-out is the point at which the inelastic collisions stop and no new species create. Then the thermal freeze-out stage begins when the elastic collisions cease and the particles freely fly to the detectors \cite{Satz03, Xu}. For more details about the freeze-out conditions and the thermal parameters characterize these stages \cite{Cleymans06, Becattini06, Andronic11}. 

The hadrons are interacting strongly and can be described by the quantum chromodynamic (QCD) \cite{Altarelli,Brambilla}. Additionally, the properties of the hot hadronic matter are best described as the properties of a statistical system, at which we can extract the thermodynamical quantities of such system depending on the Hadron Resonance Gas Model (HRG) \cite{Hagedorn65} and HRG studied at finite chemical potential \cite{Samanta, Ruiz Arriola, Hagedorn80}. As a matter of fact, the QCD can be described non-perturbativly in which the coupling constant approaches the unity \cite{Shuryak,Sterman}, this non-perturbative method is called lattice QCD (LQCD) \cite{Gupta, Sourendu, Aoki, Fodor, Kaczmarek, Bonati}

HRG is a powerful tool for the description of the thermal properties of the QCD matter and to reproduce the LQCD results at low temperature \cite{Bellwied}.  However, it is in disagreement with LQCD at high temperature, this might be a consequent of the negligence of the interaction among the particles in which the interaction is significantly effective at high temperature \cite{Bedangadas18,Vovchenko}. In addition to that, the HRG model is studied to study the effect of the magnetic field on the QCD matter in a series of papers e.g., \cite{Agasian2000, Endrodi15, Ayalaa16, Tawfik16, Braguta19}, a good review in the effect of magnetic field in vacuum \cite{Shuryak84} and also in in the a lattice QCD such as \cite{Bali12, Bali14,Buividovich}.
An external magnetic field B is generated by the spectators particles (particles do not contribute in the interaction) in the  non-central heavy-ion collision. According to the illustration of the non-central heavy ion collision the induced magnetic field can be estimated \cite{Huang, Skokov}.

Numerical estimation of the magnetic field Au-Au in (RHIC) at $\sqrt{S} = 200 \hspace{0.05cm}GeV$ Which produces a magnetic field $( \sim 10^{18}−10^{19})$ Gauss and for Pb-Pb in (LHC) at $ \sqrt{S} = 2.76 \hspace{0.08cm} TeV \sim 10^{20}$ Gauss $ \sim 10 \hspace{0.05cm} m^{2}_{\pi} \hspace{0.05cm} GeV^{2}$ \cite{adam_HI, LI_HI, Dimitri_HI} \footnote{ The pion mass $m^{2}_{\pi}$ in $GeV^{2}$ is taken as the unit of eB, where e is the electron charge and $m_{\pi} ≈ 140 \hspace{0.03cm}MeV$, where $1 MeV^{2} = e \cdot 1.6904 \times 10^{14}$ Gauss  with ($\hbar  = c = 1 $)}.
 
 \subsection{Motivation}
The effect of magnetic catalysis (MC) and $/$ or inverse magnetic catalysis (IMC) mechanism still lacks from the complete understanding in particular on the thermodynamics of the quantum chromodynamics (QCD) matter. This is the main motivation of the present work to  make more investigation in the non-extensive statistics along with the extensive one.  Beside the recent lattice results have intriguing properties and characteristics.

 The effect of magnetic field on QCD matter which is known as the magnetic catalysis (MC) in which it exhibits an increasing in the quark condensate at very low temperature much below the $T_{c}$, this phenomenon is attributed to the valence and the sea contributions to the quark condensate, therefore both contributions are increasing as a function of the magnetic field at ($T = 0$). As a matter of fact, whatever the kind of charges (valence or sea), the magnetic field affects the motion of them (catalyzing effect) and makes a restriction on the motion of the charged particles to move in a direction perpendicular to the magnetic field, in other words it makes a "dimensional reduction" \cite{Mag_Book} which is treated mathematically in various works and here in the present. However, in contrary to this, it is found that MC effect converted to inverse MC (inverse magnetic catalysis) near the critical temperature Tc and at ($ eB < 1 \hspace{0.05cm} GeV^{2}$) and especially for increasing the pion mass \cite{Endrodi19}. In the latter means that the quark condensate decreases with the magnetic field \cite{Elia18}. Lattice calculations also have further investigation for both MC and IMC \cite{Ilgenfritz13, Bornyakov14, Bali14}.

Strong magnetic field has an important role in physical systems, such as the early universe explanation, the phase transition, cosmology \cite{Bali12R}. Additionally, the magnetic field that generated in the peripheral heavy ion collisions  at the Relativistic Heavy Ion Collider (RHIC) or the Large Hadron Collider (LHC) \cite{Voronyuk, Skokov}. The analysis of the particle production at different energies are studied in \cite{Becattini04, Abhijit2020}, extensive and non-extensive is applied on black hole in which the black hole appeared as an extensive system \cite{Shalaby16}. Recently, and up to three decades the non-extensive statistics has been applied to many physical systems and showed a great success in the description of QCD matter \cite{Rozynek19}, neutron stars \cite{Parta12} and cosmology \cite{Shalaby2021}.

The work is organized as follows: \cref{sec:HRG model} represents the hadron resonance gas model and the thermodynamical quantities of the hadronic system at zero and non-zero magnetic field. SubSection \ref{free_eng}, discussed the vacuum and thermal free energy contributions. Section \ref{sec:Non-ext}, is devoted to study the non-extensive statistics in details. The results and discussion is represented in \cref{sec:Res}. Finally, the concluding remarks is introduced in \cref{sec:con}, followed by two appendices \ref{AppA}, \ref{AppB}.

\section{Hadron Resonance Gas model (HRG) }\label{sec:HRG model}
The focus of this work is to examine the extensive and non-extensive
thermodynamics in the hadronic system. Beginning with the extensive thermodynamics which can be emerged within the HRG model. The foremost known form of the entropy is the Boltzmann-Gibbs (BG) for various and discrete states $W$, as follows,
\begin{equation}\label{eq:BG_ent}
S_{BG} = -k_{B} \sum_{i = 1}^{W} P_{i} \hspace{0.1cm} ln \hspace{0.1cm}P_{i}\,,
\end{equation}
with $k_{B}$ is the Boltzmann constant, and the sum $ \sum_{i =
1}^{W} P_{i} = 1 $  takes into account all the possible microscopic configurations $W$, and the probability of each state $i$ is $P_{i}$. For the particular case $P_{i} = 1/W$
and for equal probabilities Eq. (\ref{eq:BG_ent}) can be rewritten \cite{boltzmann,gibbs},
\begin{equation}
S_{BG} = k_{B} \hspace{0.1cm} ln \hspace{0.1cm} W\,.
\end{equation}
In the thermal equilibrium, the probability of the system is defined in terms of the temperature, T  
\begin{equation}
P_{i} = \frac{e^{-\beta E_{i}}}{Z_{BG}}\,,
\end{equation}
where, $E_{i}$ is the energy of the $i^{th}$ state of the system, $\beta = \frac{1}{k_{B} T}$, and  $Z_{BG}$ is the partition function, the latter is defines as \cite{reif2009fundamentals},
 \begin{equation}
Z_{BG} = \sum_{i}^{W} e^{-\beta E_{i}}\,.
\end{equation}
The canonical partition function has been studied by different methods, e.g. Refs.\cite{Bohr69,Gupta81,Chase95,Gupta98,Keranen}. The most remarkable point in Boltzmann-Gibbs statistics is the additive property, where a system $A$ composed of two subsystems
$A_{1}$ and  $A_{2}$, so that the total entropy can be written as,
\begin{equation} \label{eq:ext_entropy}
S_{BG}(A) = S_{BG}(A_{1}) + S_{BG}(A_{2})\,.   
\end{equation}
In the grand-canonical ensemble, the partition function of an ideal gas consisting of hadrons and resonances for the i particle   is given as \cite{Abhijit} \:
\begin{equation}\label{eq:partfun}
\mathrm{\ln Z_{i} = \pm \frac{V\hspace{0.02cm}g_{i}}{(2\pi)^3}\int d^{3}p\ln \left[ 1\pm \lambda_{i}\exp\left( \frac{-E_{i}(p)}{T}\right)\right] } ,
\end{equation}
with the grand canonical partition function is just a sum over all resonances $(ln \hspace{0.03cm} Z =\sum_{i} ln\hspace{0.03cm} Z_{i})$
where $\pm$ refer to fermions and bosons, respectively, $\mathrm{g_{i}}$ is degeneracy, $\lambda_{i}$ is defined as \cite{Tawfik_Shalaby}
\begin{equation}
\lambda_{i}(\mu, T) = exp \left( \frac{ \mu_{s}S_{i} + \mu_{B}B_{i} + \mu_{q}Q_{i}}{T}\right), 
\end{equation}
where the chemical potential is defined  $\mathrm{\mu_{s}}$, $\mathrm{\mu_{B}}$, $\mathrm{\mu_{q}}$ are the strange, baryon and quark chemical potential, respectively, multiplied by corresponding quantum numbers S, B, Q,  $\mathrm{E_{i} = \sqrt{\mathrm{p^{2} + m_{i}^{2}}}}$ is the relativistic particle energy, in which ($\hbar = c=k_{B} = 1$).\\

In the present work, we are going to study the thermodynamics at zero and non-zero magnetic field in both statistics. Now  We represent the free energy formula followed by the free energy with the effect of the magnetic field in order to derive the thermodynamics of the hadronic system. First, the free energy reads \cite{endrodi}
\begin{equation}\label{eq:Free_total}
\mathrm{F = F_{vac} + F_{therm}}
\end{equation}
Where $\mathrm{F_{vac}}$ and $\mathrm{F_{therm}}$ are vacuum and thermal energies respectively. The free energy F is given in terms of the total internal energy U \cite{letessier2002hadrons},
\begin{equation}
F(V, T)= U- T\hspace{0.03cm}S
\end{equation}
Then replacing all the quantities by their corresponding densities, (e.g. $s = \frac{S}{V}, \varepsilon =\frac{U}{V}, f = \frac{F}{V}= -p$). The general form would be the Gibbs-Duham relation which is given by eq. (\ref{Gibbs-Duham}) \cite{letessier2002hadrons} in addition of the presence of the magnetic field, 

\begin{equation} \label{Gibbs-Duham}
\mathrm{\varepsilon = T \hspace{0.03cm}s + B\hspace{0.03cm}m_{B} -p}
\end{equation}
Where $( m_{B} = \frac{M_{B}}{V})$, and the magnetization $M_{B}$ is defined as \cite{Andersen14}
\begin{equation}
q\hspace{0.02cm}M_{B} = \left( - \frac{\partial F}{\partial B}\right).
\end{equation}
Once the partition partition function is known, all thermodynamical observables can be calculated e.g. the pressure, P, number density, n, energy density, $\varepsilon$, and entropy for particle \textbf{i} \cite{letessier2002hadrons}. Moreover, the free energy is directly related to the partition function so that the thermodynamical quantities can be got from the free energy too as $F_{thermal}(V,T) = -\beta^{-1} ln \hspace{0.03cm}Z(V,\beta)$:
\begin{equation}\label{eq:all_thermo_derv}
p = \left( \frac{-\partial F}{\partial V}\right) , \quad N =  \left( \frac{-\partial F}{\partial \mu}\right) , \quad  S = \left( -\frac{\partial F}{\partial T}\right)
\end{equation}
 It is known that, the partition function $ln\hspace{0.03cm} Z$ and $F$ are extensive quantities which leads to that, the derivative with respect the V converts to ($\frac{1}{V}$) this satisfied at the thermodynamic limit (i.e at $V \longrightarrow\infty$).\\
\subsection{The Free energy density contributions} \label{free_eng}
 In this part we introduce the free energy different contributions according to eq.(\ref{eq:Free_total}) at zero and non-zero magnetic field, in other words for the neutral and charged particles respectively. 
\begin{itemize}
 \item  Firstly, Free energy density at zero magnetic field  $B = 0$ \cite{endrodi,Kapusta06}. 
 
\begin{equation}\label{eq:free_zeroBAll_res}
f(s) = \mp \sum_{i} \sum_{s_{z}} \int \frac{d^{3}\textbf{p}}{(2 \pi)^{3}}\left( \frac{E_{i}(\textbf{p})}{2} + T\log \left[ 1\pm exp\left(\frac{\mu_{i}-E_{i}(\textbf{p})}{T}\right) \right]  \right); \quad \hspace{0.03cm}  q_{i} = 0, q_{i}\neq 0 
\end{equation}

Again, the relativistic energy for particle i in three-dimension momentum $\textbf{p}$ ,
 \begin{equation}
E_{i}(\textbf{p}) = \sqrt{\textbf{p}^{2} + m_{i}^{2}}
\end{equation}

Where $ E_{\textbf{p}} $ is the relativistic energy for neutral and charged particles, in units ($\hbar = c = k_{B} = 1$). 
\item  Free energy density at non-zero magnetic field $ B\neq 0$ \cite{endrodi,Fraga08}. \\
 \end{itemize} 
 The free energy at the presence of the magnetic field ($B \neq 0$). 

\begin{equation}\label{eq:free_nonzeroBAll_res}
f_{charge}(s) = \mp \sum_{i} \sum_{s_{z}} \sum_{k=0}^{\infty} \frac{|q_{i}|B}{(2 \pi)^{2}} \int dp_{z}\left( \frac{E_{i}(p_{z},k,s_{z})}{2} + T\log \left[ 1\pm \exp \left( \frac{\mu_{i}-E_{i}(p_{z},k,s_{z})}{T}\right) \right] \right); q_{i}\neq 0
\end{equation} 
Where the integral over $d^{3}p$ in eq. (\ref{eq:partfun}) is polarized in z-direction for $i^{th}$ particle,

\begin{equation}\label{eq:phase_conv}
\mathrm{\int d^{3}p \rightarrow 2\pi|Q_{i}|eB_{z}\sum_{k}\sum_{s_{z}}\int dp_{z}}
\end{equation}

 In comparison with eq. (\ref{eq:Free_total}), the first part is the contribution due to vacuum and the second part due to the thermal contribution. Where $\mp$ corresponds to bosons (lower sign) and to fermions (upper sign) which in turn due to the spin sectors bosons with integer spin and for fermions is half-integer. The spin takes the values, $(\mathrm{s_{z} = -s,..,s})$ is the z-component of the particle spin (s). Where the modified energy due to Landau levels, k, is defined as \cite{Landau77},
 \begin{equation}\label{eq:polarized_eng}
E_{i_{z}}(p_{z}, k, s_{z}) = \sqrt{p_{z}^{2} + m_{i}^{2} + 2|q_{i}|\hspace{0.03cm}B\hspace{0.03cm}\left( k-s_{z}+ \frac{1}{2}\right) }
\end{equation}

 With $q_{i} = Q_{i} \hspace{0.02cm}e $ is the charge of the particle i, mass $m_{i}$, and the electron charge e.

\footnote{In eqs. (\ref{eq:free_nonzeroBAll_res}, \ref{eq:free_zeroBAll_res}) we add another summation over all the hadronic particles i, and also we can add the chemical potential in the exponential of the thermal part}. In the present work we performed the numerical calculations for Particle Data Group (PDG) with masses up to 10 GeV \cite{PDG14}.

It is found that the vacuum parts in both eqs. (\ref{eq:free_nonzeroBAll_res}, \ref{eq:free_zeroBAll_res}) are ultraviolet divergent and need to be normalized  through dimensional regularization method \cite{ George75}, and we followed the steps the details in \cite{endrodi} to get the following:\\
As mentioned above, the free energy is composed of two parts, vacuum part and thermal part. We are going to define each part at zero and non-zero magnetic field. Firstly, at $B = 0$ the vacuum part is defined as, \\
\begin{equation}\label{eq:Norm_freeZeroB}
f_{vac}(s,B = 0) = \pm(2s+1)\frac{(q\hspace{0.03cm}B)^{2}}{8 \pi^{2}}\hspace{0.03cm}v^{2} \left[ \frac{1}{\epsilon}+\frac{3}{4}-\frac{\gamma}{2}-\frac{1}{2}\log\left( \frac{2q\hspace{0.03cm}B}{4\pi \alpha^{2}}\right) -\frac{1}{2}\log(v)\right]
\end{equation}
with parameter $\epsilon$ and scale $\alpha$ and Euler Lagrange $\gamma$, where $v = \frac{m^{2}}{2q\hspace{0.03cm}B}$ removes the dependence on B itself.

And the normalized free energy vacuum part at $ B \neq 0$ is defined as,\\
\begin{equation}\label{eq:Norm_freenonZeroB}
f_{vac}(s,B \neq 0) = \pm \frac{(q\hspace{0.03cm}B)^{2}}{8\pi^{2}} \sum_{a}\left[ \left( -\frac{2}{\epsilon}+\gamma+\log\left( \frac{2q\hspace{0.03cm}B}{4\pi \alpha^{2}}\right) - 1\right) \left( -\frac{1}{12} - \frac{(v+a)^{2}}{2} + \frac{v+a}{2}\right) - \zeta^{\backprime}(-1,v+a)\right] 
\end{equation}

where  $\zeta$ is the Hurwitz function (see \ref{AppA}) emerged from the conversion of the sum over k, and $ a = 1/2 +s_{z} $.\\
The change in the free energy density due to the magnetic field, one subtract the B = 0 eq.(\ref{eq:Norm_freeZeroB}). All details of the normalization see refs.\cite{endrodi, Menezes08}, we write down the final normalized free energies as obtained in \cite{endrodi}. In the present work we have performed the calculation for different sectors of the spin of particles (s $=$ 0, 1/2 and 1). Consequently, the renormalized free energy for those sectors of spins are defined as:

\begin{equation} \label{eq:Norm_nonZeroB0}
\Delta f^{vac}(0) = \frac{(q\hspace{0.02cm}B)^{2}}{8\pi^{2}} \left[ \zeta^{\backprime}(-1,v+\frac{1}{2})+\frac{v^{2}}{4} - \frac{v^{2}}{2}\hspace{0.02cm}log(v)+ \frac{log(v)+1}{24}\right] 
\end{equation}
\begin{equation}\label{eq:Norm_nonZeroB0.5}
\Delta f^{vac}(1/2) = -\frac{(q\hspace{0.02cm}B)^{2}}{4\pi^{2}} \left[ \zeta^{\backprime}(-1,v)+\frac{v}{2}log(v) + \frac{v^{2}}{4} - \frac{v^{2}}{2}log(v)- \frac{log(v)+1}{12}\right] 
\end{equation}

\begin{equation}\label{eq:Norm_nonZeroB1}
\Delta f^{vac}(1) = \frac{3(q\hspace{0.02cm}B)^{2}}{8\pi^{2}} \Bigg[ \zeta^{\backprime}(-1,v-\frac{1}{2})+ \frac{1}{3}(v+\frac{1}{2})\hspace{0.02cm}log(v+\frac{1}{2})+\frac{2}{3}(v-\frac{1}{2})\hspace{0.02cm}log(v-\frac{1}{2})+\frac{v^{2}}{2}\left( \frac{1}{2} - log(v)\right) - 7\frac{log(v)+1}{24}\Bigg] 
\end{equation}

\section{Essential Features of Non-extensive Statistics}\label{sec:Non-ext}
In this section we represent the non-extensive thermodynamics which is calculated at vanishing and non-zero magnetic field.
The great success of the well-established BG statistics for more than a century ago is the fact that it can be generalized for complex systems \cite{GellMann95}. For complex systems, it is not so easy to extract their properties easily, since there are time-dependent interactions among their numerous constituents. Complex systems are not used only in the domain of physics but also in other domains like biology, economics, and so on. \cite{Holovatch17}. For more details about the non-extensive statistics application in various fields in physics , cosmology and astronomy see \cite{Tsallis09, Tsallis11, Plastino93, Plastino99, Megias14, Shalaby2021}. 

Specifically, many systems deviate from the BG statistics, and it was necessary to generalize the original ones. These systems are, for example, systems with long-range interactions, ferromagnetism related systems pure-electron plasma two-dimensional turbulence systems. Complex systems effects can be found in the peculiar velocities of galaxies, black holes, high energy collisions of elementary particles, quantum entanglement, see, for example, Ref. \cite{Abe2001} and references therein.

In 1988, Tsallis suggested the non-extensive \cite{Tsallis88} in other words the non-additive property of entropy eq. (\ref{eq:ext_entropy})is no longer satisfied. Therefore the non-extensive entropy is represented as \cite{Tsallis09},

\begin{equation}\label{eq:nonext_entropy}
S_{q}(A_{1}+ A_{2}) = S_{q}(A_{1}) + S_{q}(A_{2}) + (1-q)\hspace{0.04cm}S_{q}(A_{1})S_{q}(A_{2})
\end{equation}

Where q is the non-extensive parameter or the entropic index, in the limit ($q\rightarrow 1$), one recovers the BG-statistics. Additionally, q should be greater than one, the limit of q is discussed in \cite{Umarov, Rath19, Tsallis02, TsallisMath}. 

The probability and the partition function can be re-defined,
  \begin{equation}
  P_{i}^{q} = \frac{[1-\beta (q-1) E_{i}]^{1/(q-1)}}{Z_{q}} \,, \qquad {\textrm where} \qquad   Z_{q} = \sum_{i = 1}^{W}[1-\beta (q-1) E_{i}]^{1/(q-1)}\,.
  \end{equation}
  Now we turn to define the grand canonical partition function for the hadronic system based on the non-extensive statistics. Starting with the following  definitions;

\begin{itemize}
\item The q-exponential \\
\begin{equation}  \label{eq:exp_nonext}
\begin{split}
e_{q}^{(+)}(x) = [1+(q-1)x]^\frac{1}{q-1}, \qquad x\geqslant0 \\
e_{q}^{(-)}(x) = [1+(1-q)x]^\frac{1}{1-q}, \qquad  x<0
\end{split}
\end{equation}

\item The q-logarithmic \\
\begin{equation}      \label{eq:log_nonext}
\begin{split}
\log_{q}^{(+)}(x) = \frac{x^{q-1} -1}{q-1},  \\
\log_{q}^{(-)}(x) = \frac{x^{1-q} -1}{1-q},
\end{split}
\end{equation}
\end{itemize}

Then the grand-canonical partition function is defined \cite{Megias14, Megias15}.

\begin{equation}\label{eq:nonext_partF}
\log \Xi_q(V,T,\mu_{i}) = -\xi V \int\frac{d^{3}p}{(2\pi)^3}\sum_{r=\pm}\Theta(rx) \log_{q}^{(-r)} \left(\dfrac{e_{q}^{(r)}(x)-\xi}{e_{q}^{(r)}(x)}\right)
\end{equation}

Where $x = \beta(E - \mu)$, with the particle energy of mass m is given by $E = \sqrt{p^{2}+m^{2}}$, and $\mu$ is the chemical potential. The step function  $\Theta(rx)$ within $r=>- $ applies for fermions only which is satisfied when $ \mu \leq m$, and  $\xi=\pm1$ refer to bosons and fermions respectively. \\ 
The thermodynamical non-extensive quantities can all be derived as in eq.(\ref{eq:all_thermo_derv}) by replacing $ln Z$ by the non-extensive partition function for a hadronic system eq.(\ref{eq:nonext_partF}).

Then, the pressure, number density, energy density and entropy are , eqs.(\ref{eq:App_press}, \ref{eq:App_number}, \ref{eq:App_energy}, \ref{eq:App_entropy}). :
\begin{equation} \label{eq:App_press}
P_i =  -\xi T \int\frac{d^{3}p}{(2\pi)^3}\sum_{r=\pm}\Theta(rx) \log_{q}^{(-r)} \left(\frac{e_{q}^{(r)}(x)-\xi}{e_{q}^{(r)}(x)}\right)
\end{equation}
\begin{equation} \label{eq:App_number}
n_i =  \left[ C_{N,q}(\mu,\beta,m) + \int\frac{d^{3}p}{(2\pi)^3}\sum_{r=\pm}\Theta(rx) \left(\frac{1}{e_{q}^{(r)}(x)-\xi} \right)^{\tilde{q}}\right]  
\end{equation}

Where\\
\begin{equation}  \label{eq:term_C}
C_{N,q}(\mu,\beta,m) = \frac{1}{2 \pi^{2}} \frac{\mu \sqrt{\mu^{2} - m^{2}}}{\beta} \frac{2^{q-1} + 2^{1-q} - 2}{q-1} \Theta(\mu - m)
\end{equation}
Where the part in eq.(\ref{eq:term_C}) emerged from the fact that the integral in eq.(\ref{eq:nonext_partF}) is discontinuous at $x = 0$, this leads to that the momentum $ p = \sqrt{ \mu^{2}-m^{2}}$ see details in Appendix \ref{AppB}

\begin{eqnarray}
\varepsilon &= & -\frac{\partial}{\partial \beta} \log \Xi_q\Bigg|_{\mu} + \mu \hspace{0.03cm} T \hspace{0.03cm} \frac{\partial}{\partial \mu}\log \Xi_q \Bigg|_{\beta}
\\&=& \left[  C_{E,q}(\mu,\beta,m) + \int\frac{d^{3}p}{(2\pi)^3}\sum_{r=\pm}\Theta(rx) E \left(\frac{1}{e_{q}^{(r)}(x)-\xi} \right)^{\tilde{q}}\right],\, \,. \nonumber
\end{eqnarray} \label{eq:App_energy}
where $   C_{E,q}(\mu,\beta,m) = \mu C_{N,q}(\mu,\beta,m)$
 
 \begin{eqnarray} \label{eq:App_entropy}
s &=& -\beta^{2} \hspace{0.04cm} \frac{\partial}{\partial \beta} \left(\frac{\log \Xi_q}{\beta} \right) \\ &=&   \int \frac{d^{3}p}{(2\pi)^{3}} \sum_{r=\pm} \Theta(rx) \left[ -[\overline{n}_{q}^{(r)}(x)]^{\tilde{q}} \log_{q}^{(-r)} \left( \overline{n}_{q}^{(r)}(x)\right) + \xi [1 + \xi \overline{n}_{q}^{(r)}(x)]^{\tilde{q}} \log_{q}^{(-r)} \left( 1+\xi \overline{n}_{q}^{(r)}(x) \right) \right] ,\, \,. \nonumber
\end{eqnarray}
Where $\,$ $\mathrm{\overline{n}_{q}^{(r)}(x) \equiv \left[ n_{q}^{(r)}(x) \right]^{1/ \tilde{q}}}$

\begin{equation}
\mathrm{n_{q}^{(r)}(x) = \left(\frac{1}{e_{q}^{(r)}(x)-\xi} \right)^{\tilde{q}}, \qquad with \qquad \tilde{q}}  =\left\lbrace 
\begin{array}{ll}
\mathrm{q  , \qquad x\geqslant0}\\
\mathrm{2-q , \qquad x<0}
\end{array}
\right. 
\end{equation}

The next step is to modify the non-extensive partition function due to the magnetic field and all steps of the thermodynamics at non-zero magnetic field is repeated. 
\begin{equation}
\log \Xi^{B_{z}}_q(V,T,\mu) = \frac{-\xi V}{2\pi^2} |Q|eB_{z}\sum_{k}\sum_{s_{z}}\int_{0}^{\infty} dp_{z} \sum_{r=\pm}\Theta(rx) \log_{q}^{(-r)} \left(\frac{e_{q}^{(r)}(x)-\xi}{e_{q}^{(r)}(x)}\right)
\end{equation}
following the conversion in eq.(\ref{eq:phase_conv}) and the energy in z-direction in eq.(\ref{eq:polarized_eng}).

\section{Results and discussion} \label{sec:Res}
We represent the results of two different thermodynamic statistics, extensive and non-extensive thermodynamics. The impact of the magnetic field is studied for both statistics, in which the thermodynamical observables such as,( pressure, energy density, entropy and magnetization) are investigated. Both statistics are studied at zero and non-zero magnetic field. The response of the QCD matter to the magnetic field is examined through the magnetization which exhibits as paramagnetic material. All our results are studied at vanishing baryon chemical potential. 
\subsection{Extensive Individual Particle Contribution and Thermodynamics}\label{subsec:indv1}
\begin{figure}
\centering
\subfigure[~The (colored online) pressure for individual particles at vanishing magnetic field $ B = 0$ ]{\label{fig:1a} \includegraphics[width=10.cm, height=7.cm]{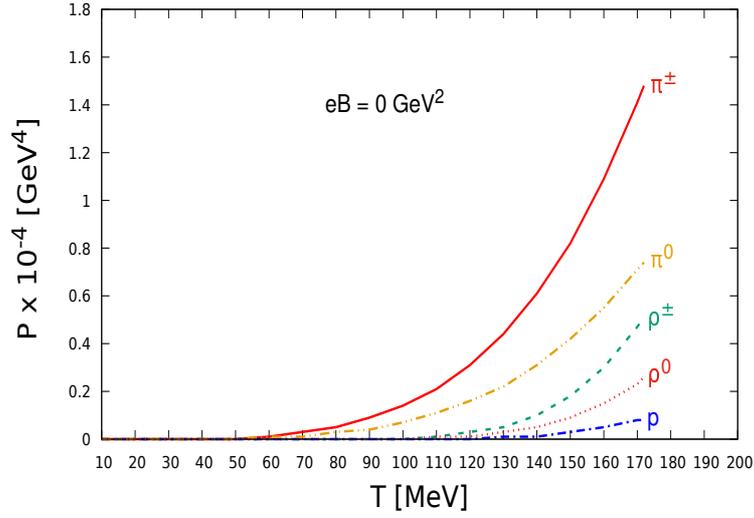}}
\subfigure[~The (colored online) pressure for individual particles at non-zero magnetic field$B \neq 0$]{\label{fig:1b} \includegraphics[width=10.cm, height=7.cm]{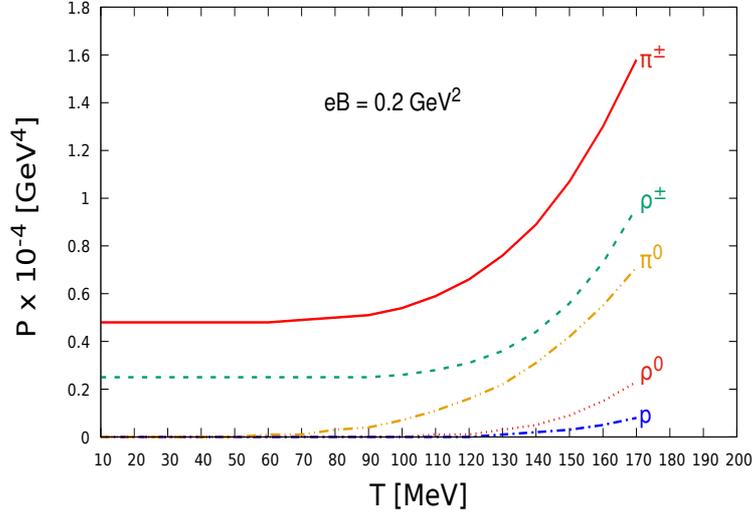}}
\caption[figtopcap]{ The pressure for particles at zero and non-zero magnetic field calculated for p (blue), $\rho^{0,\pm}$ (dotted red and green), $\pi^{0,\pm}$ (yellow and solid red) in left and right panels.} \label{fig:individual_contr}
\end{figure}
\newpage

Figure (\ref{fig:individual_contr}) shows the pressure for the particles $p, \rho^{0,\pm}, \pi^{0,\pm}$ the left panel (\ref{fig:1a}) for vanishing magnetic field and the right panel (\ref{fig:1b}) for  magnetic field $( eB = 0.2 \hspace{0.04cm} GeV^{2}) $ versus the temperature T. It is noticed that, the charged particles pressure differs from zero magnetic field to the one at finite magnetic field. This difference appears at temperature ($ 100 < T < 140 \hspace{0.04cm}MeV$), in which $\rho^{\pm}$ shifted by $\sim0.2 \times 10^{-4} \hspace{0.04cm}GeV^{4}$, and $\pi^{\pm}$ shifted by $\sim0.5 \times 10^{-4} \hspace{0.04cm}GeV^{4}$. The domination of the charged pions is clearly appeared in both cases. It is noticeable that, the pressure for the charged particles are shifted because of the vacuum term included and confirms the effect of the magnetic field only on the charged hadrons.

\begin{figure}[H]
  \centering 
      \setlength\abovecaptionskip{-0.05\baselineskip} 
  \subfigure[ ]{\label{fig:press_ext}\includegraphics[height=7cm, width=10cm]{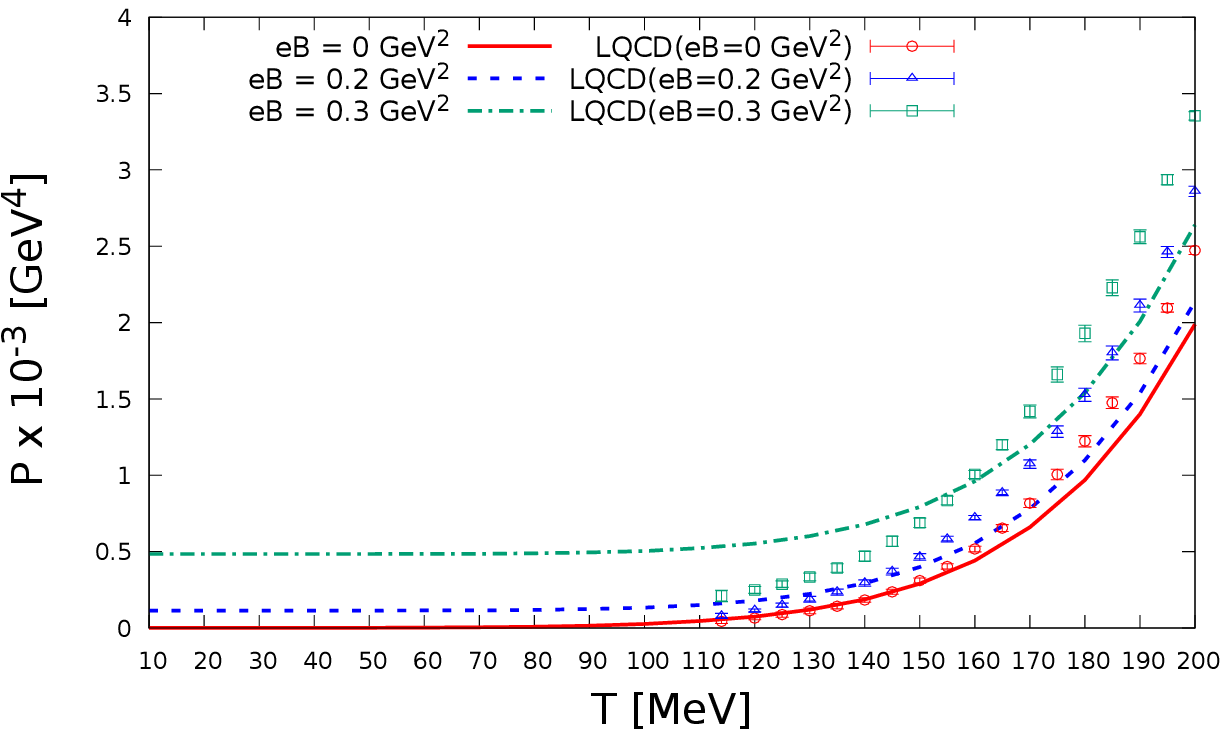}}\qquad
  \subfigure[]{\label{fig:eng_ext}\includegraphics[height=7cm, width=10cm]{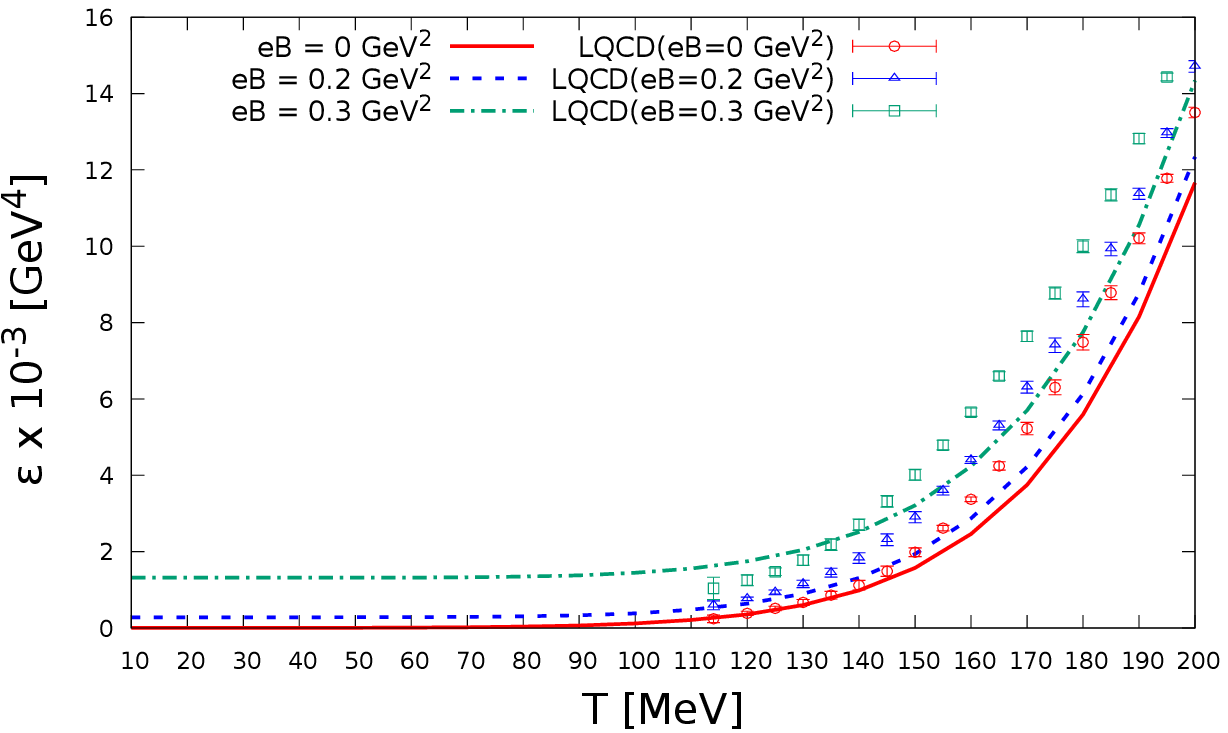}}
   \caption[figtopcap]{The Colored-online pressure (upper panel) versus the temperature \subref{fig:press_ext}, with (red-line),(dashed-blue) and (dotted-dashed green) at ($eB = 0, 0.2, 0.3 \hspace{0.07cm} GeV^{2}$) respectively compared with the corresponding lattice data (colored-online symbols).  The energy density (lower-panel) at the same conditions  \subref{fig:eng_ext}.}
  \label{fig:press_eng}
\end{figure}

\begin{figure}[H]
  \centering 
      \setlength\abovecaptionskip{-0.05\baselineskip} 
   \subfigure[ ]{\label{fig:ent_ext11}\includegraphics[height=7cm, width=10cm]{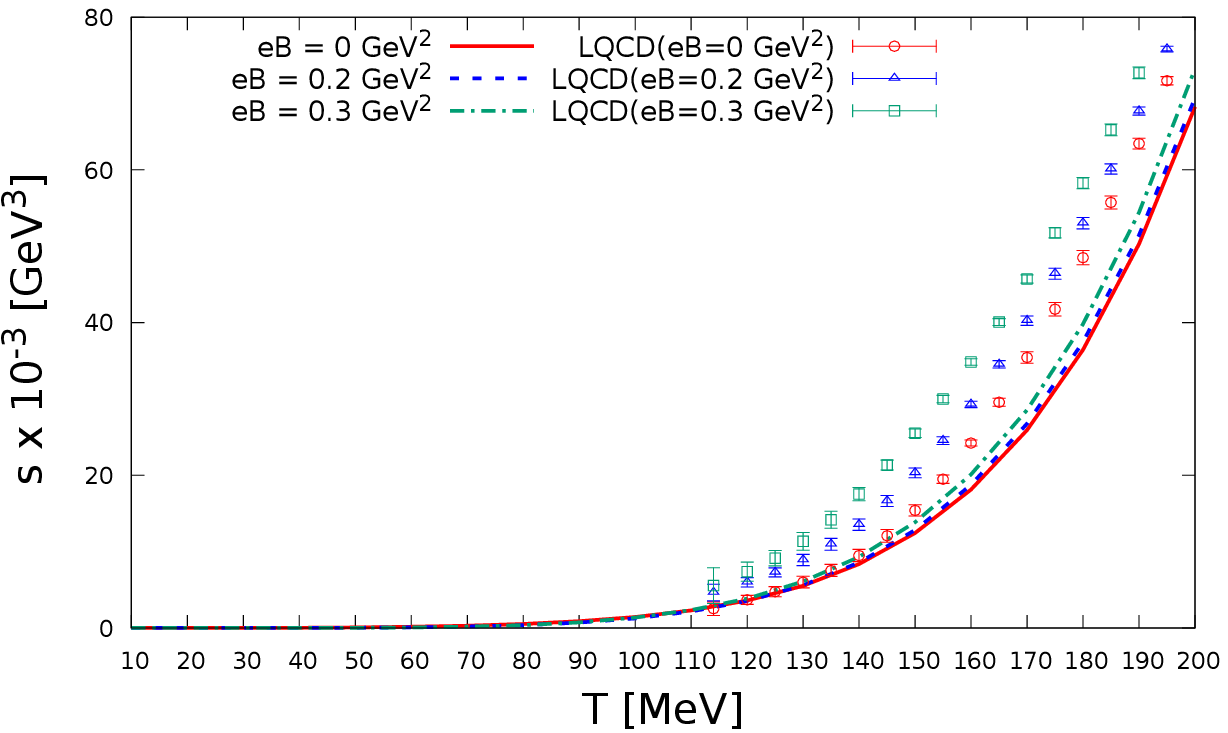}}
   \subfigure[]{\label{fig:mag_tot11}\includegraphics[height=7cm, width=10cm]{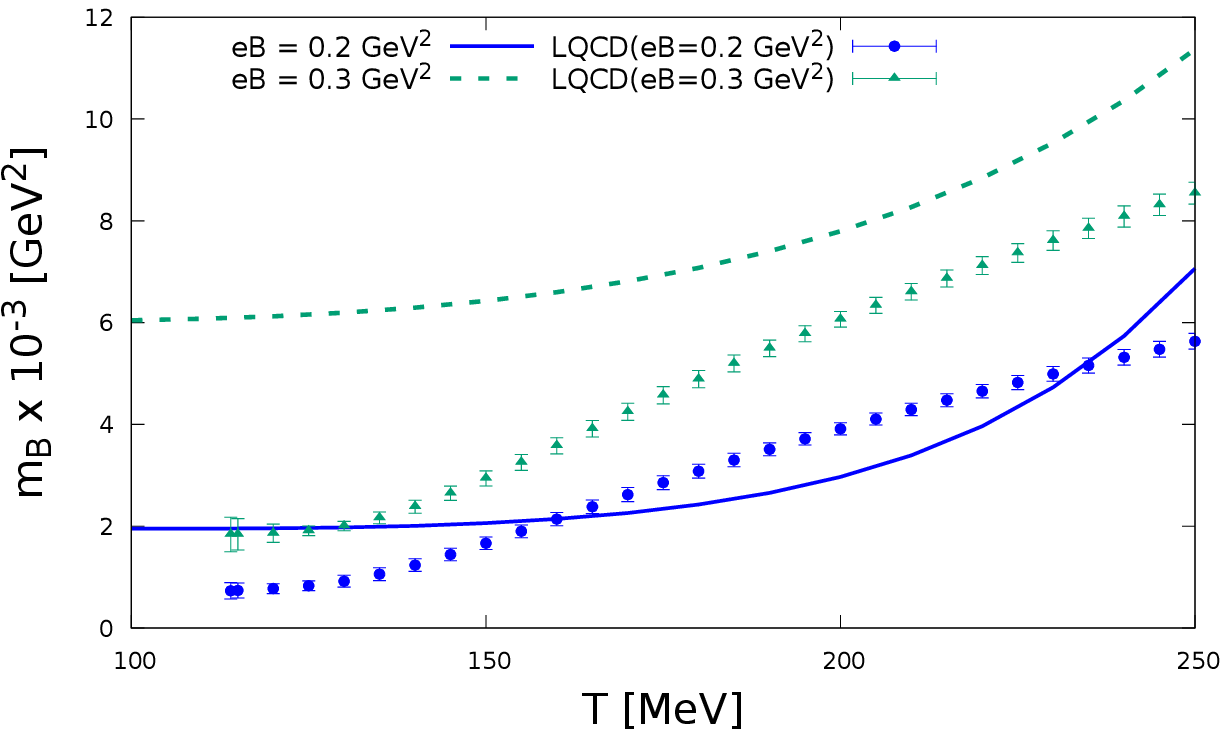}}
      \caption[figtopcap]{The Colored-online  entropy (upper panel) \subref{fig:ent_ext11} versus the temperature with (red-line),(dashed-blue) and (dotted-dashed green) at ($eB = 0, 0.2, 0.3 \hspace{0.07cm} GeV^{2}$) respectively compared with the corresponding lattice data (colored-online symbols), and finally the magnetization \subref{fig:mag_tot11} (lower-panel).}
  \label{fig:press_eng11}
\end{figure}

 Figure (\ref{fig:press_eng}) represents the EoS of QCD matter at ($eB = 0, 0.2, 0.3 \hspace{0.07cm} GeV^{2}$). The pressure fig.(\ref{fig:press_ext}) versus the temperature shows a reasonable agreement with the lattice results \cite{Bali14}, the pressure also exhibits an increasing by increasing the magnetic field. Figure (\ref{fig:eng_ext} represents the energy density, one can see the increasing in the energy density with increasing the magnetic field. The entropy and  magnetization of the QCD matter fig. (\ref{fig:press_eng11}) are calculated at ($ eB = 0.2, 0.3 \hspace{0.05cm} GeV^{2}$), it is remarkable that, the magnetization increases with increasing the magnetic field as appeared in (\ref{fig:mag_tot11}), while the increasing in the entropy is slight as shown in fig.\ref{fig:ent_ext11}).  
 

\subsection{ Non-Extensive Individual Particle Contribution and thermodynamics} \label{subsec:nonext}
\begin{figure}
\centering
\subfigure[~The (colored online) pressure for individual particles at vanishing magnetic field $ B = 0$ ]{\label{fig:1_a} \includegraphics[width=10.cm, height=7.cm]{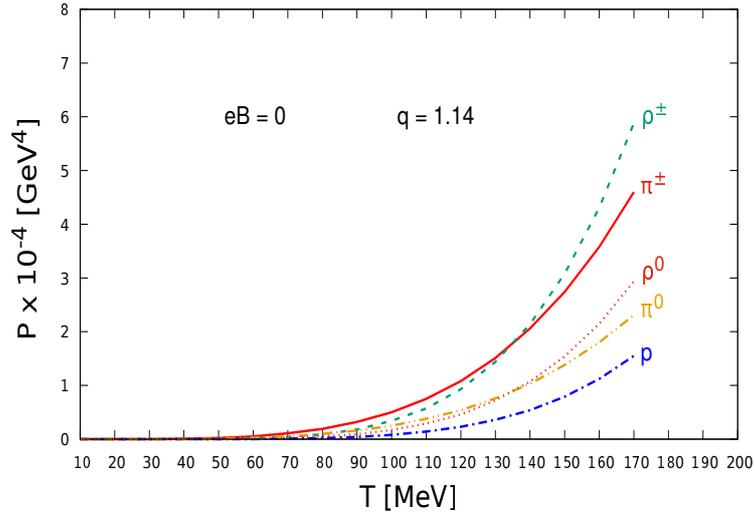}}
\subfigure[~The (colored online) pressure for individual particles at non-zero magnetic field$B \neq 0$]{\label{fig:1_b} \includegraphics[width=10.cm, height=7.cm]{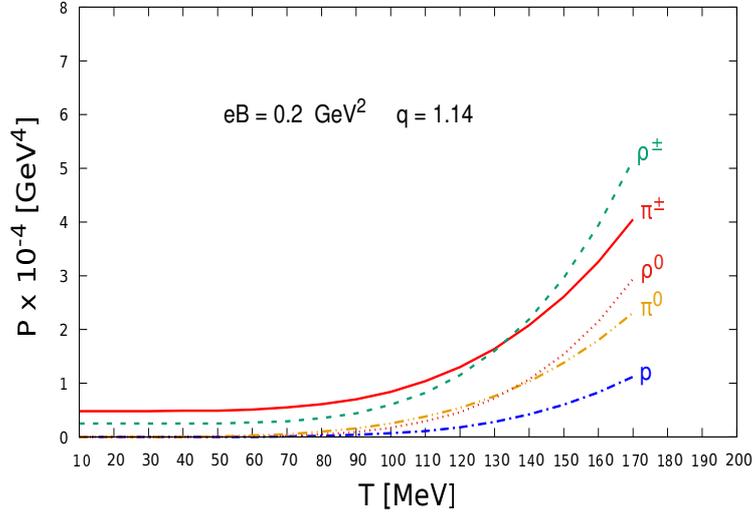}}
\caption[figtopcap]{ The pressure for particles at zero and non-zero magnetic field calculated for p (blue), $\rho^{0,\pm}$ (dotted red and green), $\pi^{0,\pm}$ (yellow and solid red) in upper and lower panels.} \label{fig:individual_contr1}
\end{figure}

\begin{figure}[H]
  \centering 
  \subfigure[]{\label{fig:press_non}\includegraphics[height=7cm, width=10cm]{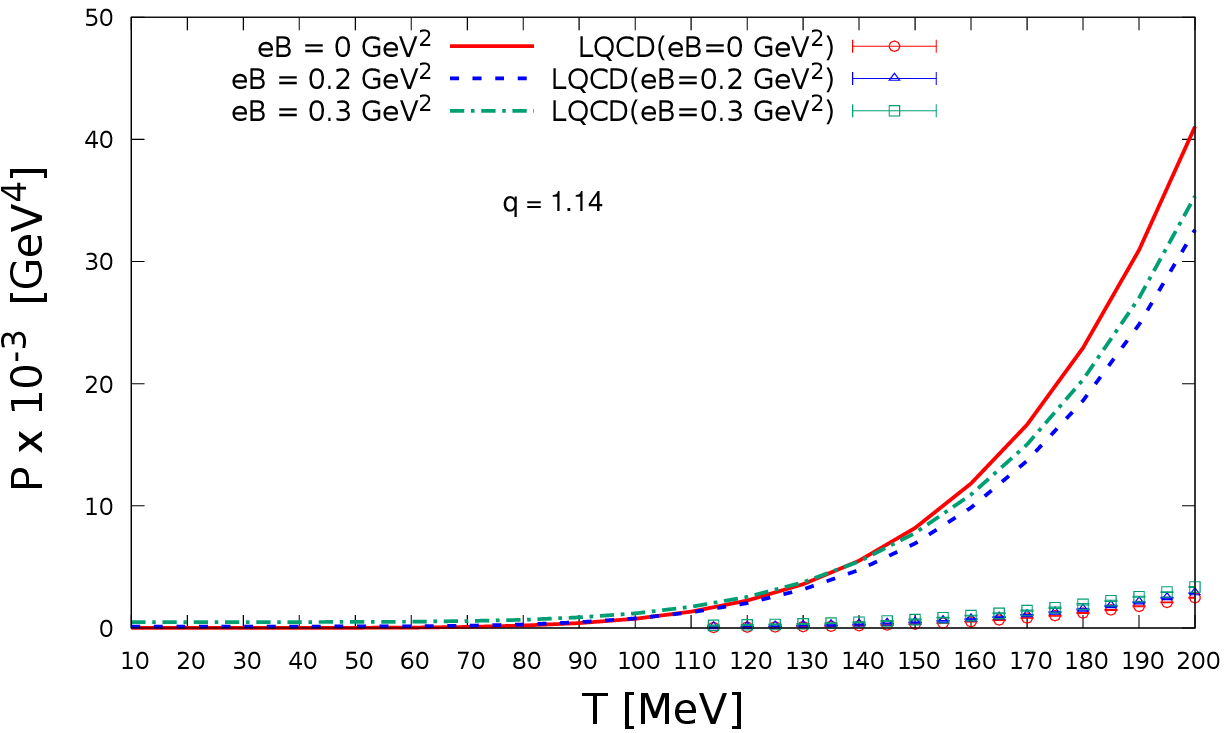}}\qquad
  \subfigure[]{\label{fig:eng_non}\includegraphics[height=7cm, width=10cm]{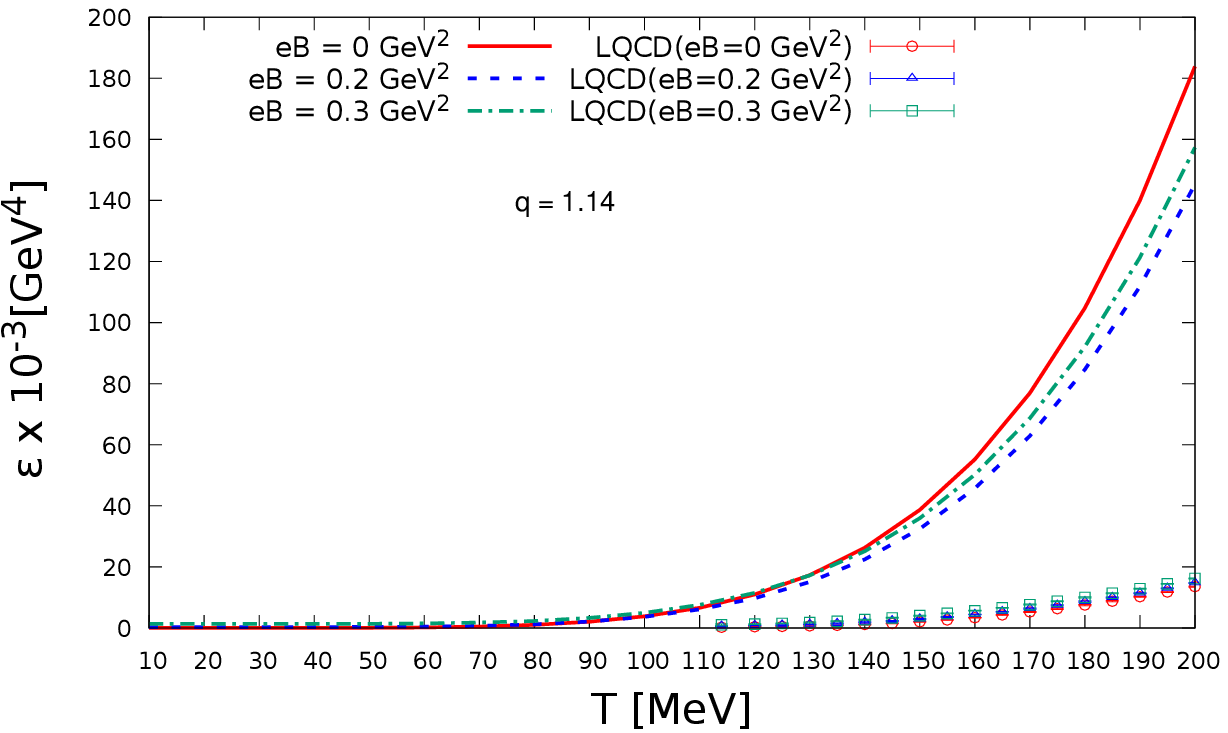}}
       \caption[figtopcap]{The Colored-online pressure (upper panel) versus the temperature \subref{fig:press_non}, with (red-line),(dashed-blue) and (dotted-dashed green) at ($eB = 0, 0.2, 0.3 \hspace{0.07cm} GeV^{2}$) respectively compared with the corresponding lattice data (colored-online symbols).  (lower -panel) the energy density at the same conditions  \subref{fig:eng_non}.}
  \label{fig:nonpress_eng}
\end{figure}

\begin{figure}[H]
  \centering 
    \subfigure[]{\label{fig:ent_non}\includegraphics[height=7cm, width=10cm]{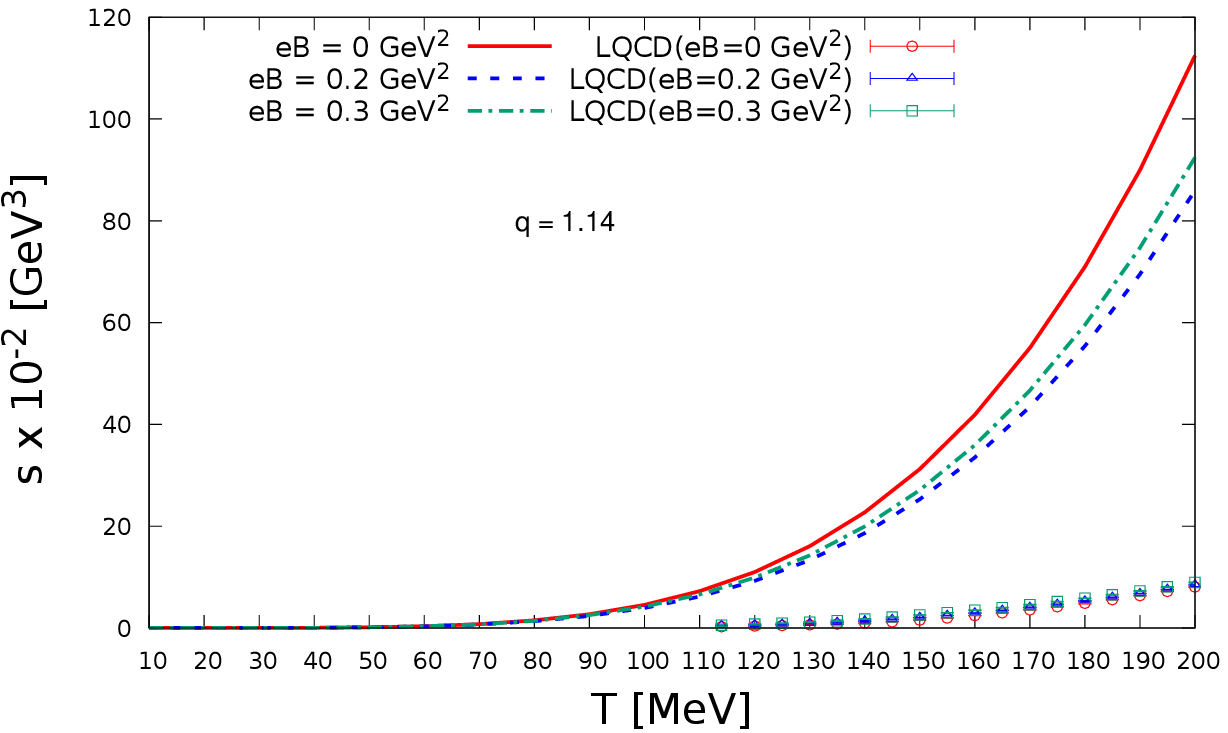}}
        \subfigure[]{\label{fig:mag_non}\includegraphics[height=7cm, width=10.cm]{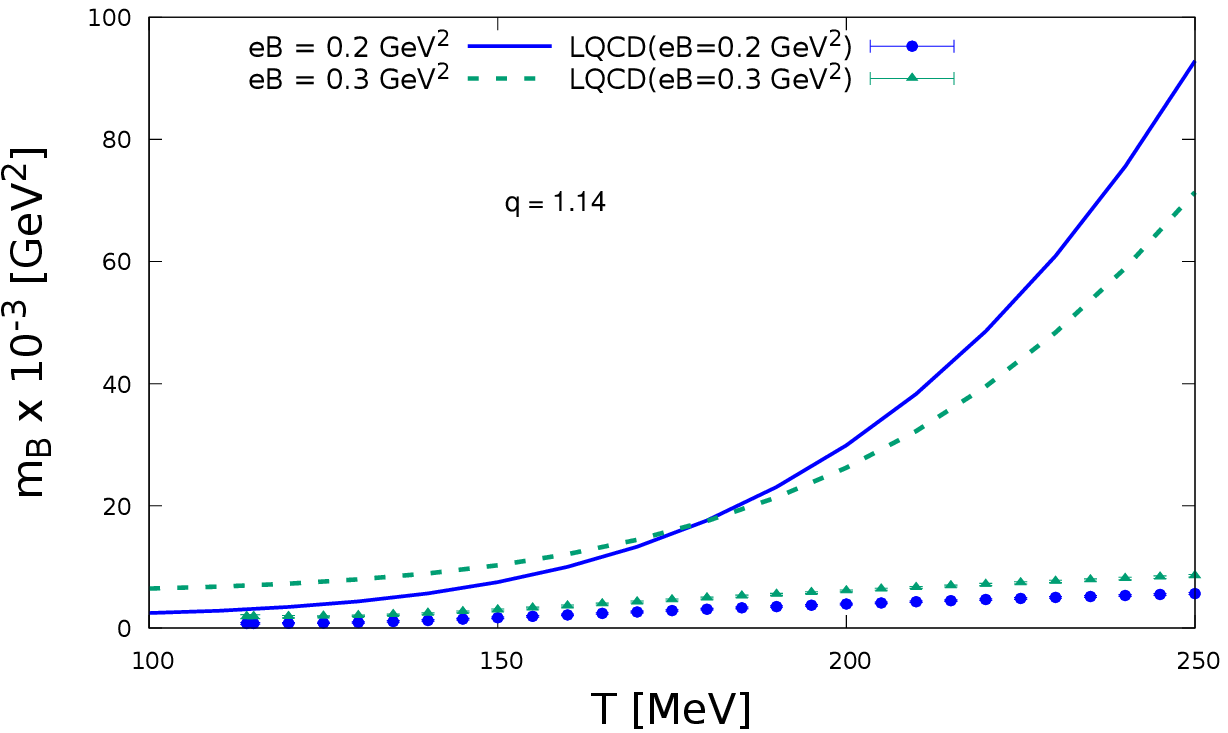}}
      \caption[figtopcap]{The Colored-online The entropy (left-upper panel) versus the temperature \subref{fig:ent_non}, with (red-line),(dashed-blue) and (dotted-dashed green) at ($eB = 0, 0.2, 0.3 \hspace{0.07cm} GeV^{2}$) respectively compared with the corresponding lattice data (colored-online symbols). The magnetization \subref{fig:mag_non} (lower -panel).}
        \label{fig:nonpress_eng1}
\end{figure}

In a similar case the thermodynamical quantities are studied by the non-extensive statistics with zero and non-zero magnetic field.
The individual pressure of charged and neutral particles are represented in fig. (\ref{fig:individual_contr1}). This figure is calculated by the non-extensive statistics, one can see a similar results for the one in fig.(\ref{fig:individual_contr}), and the effect of vacuum term that appeared at $T \geq 0$ or at very low temperature. This is obvious in the shifting of charged pressure at $B \neq 0$.
Accordingly, Fig.(\ref{fig:nonpress_eng}) pressure (\ref{fig:press_non}), energy density \ref{fig:eng_non}, and fig. (\ref{fig:nonpress_eng1}) in which entropy \ref{fig:ent_non}, and magnetization \ref{fig:mag_non}, all the results are confronted to the lattice results. It is surprisingly noticed at $(eB = 0 \hspace{0.05cm}GeV^{2})$ that, the thermodynamics overestimated the one at $(eB = 0.2, 0.3\hspace{0.05cm}GeV^{2})$ in particular at higher temperature ($100 < T < 200 \hspace{0.05cm}MeV$).\\

\newpage
\begin{figure}[H]
  \centering 
  \subfigure[]{\label{fig:dens1}\includegraphics[height=5cm, width=7cm]{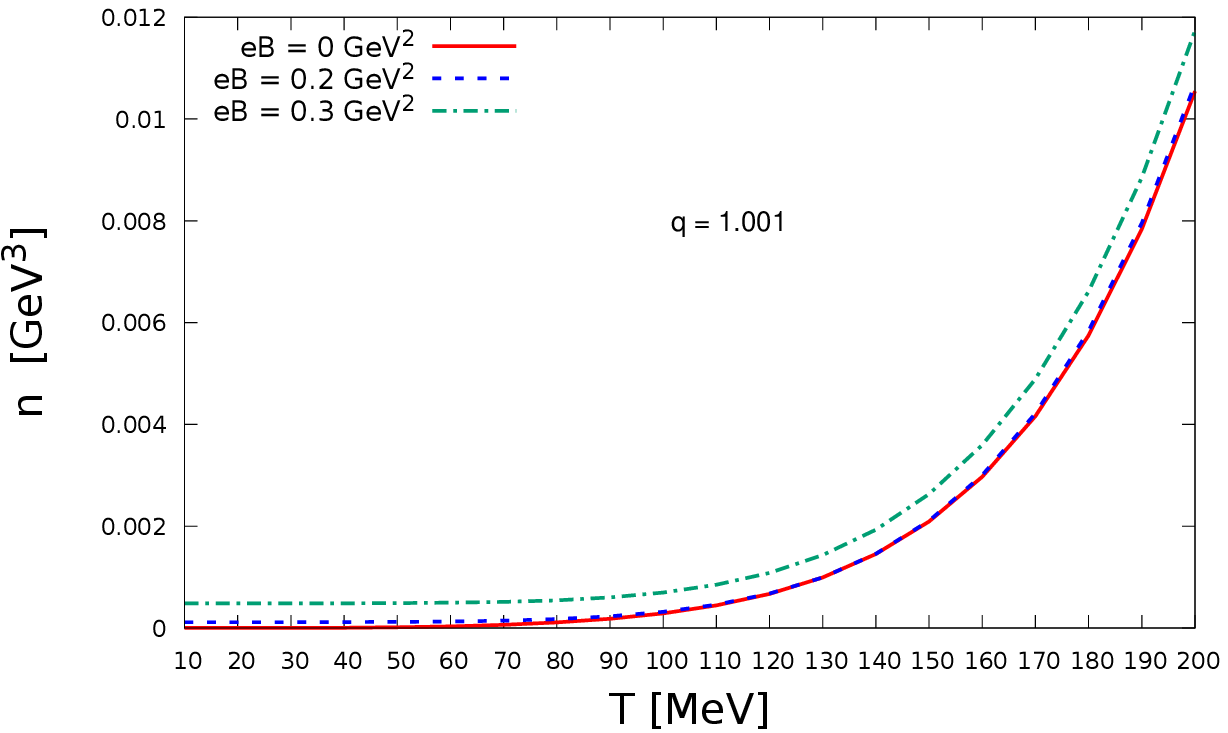}}\qquad
  \subfigure[]{\label{fig:dens2}\includegraphics[height=5cm, width=7cm]{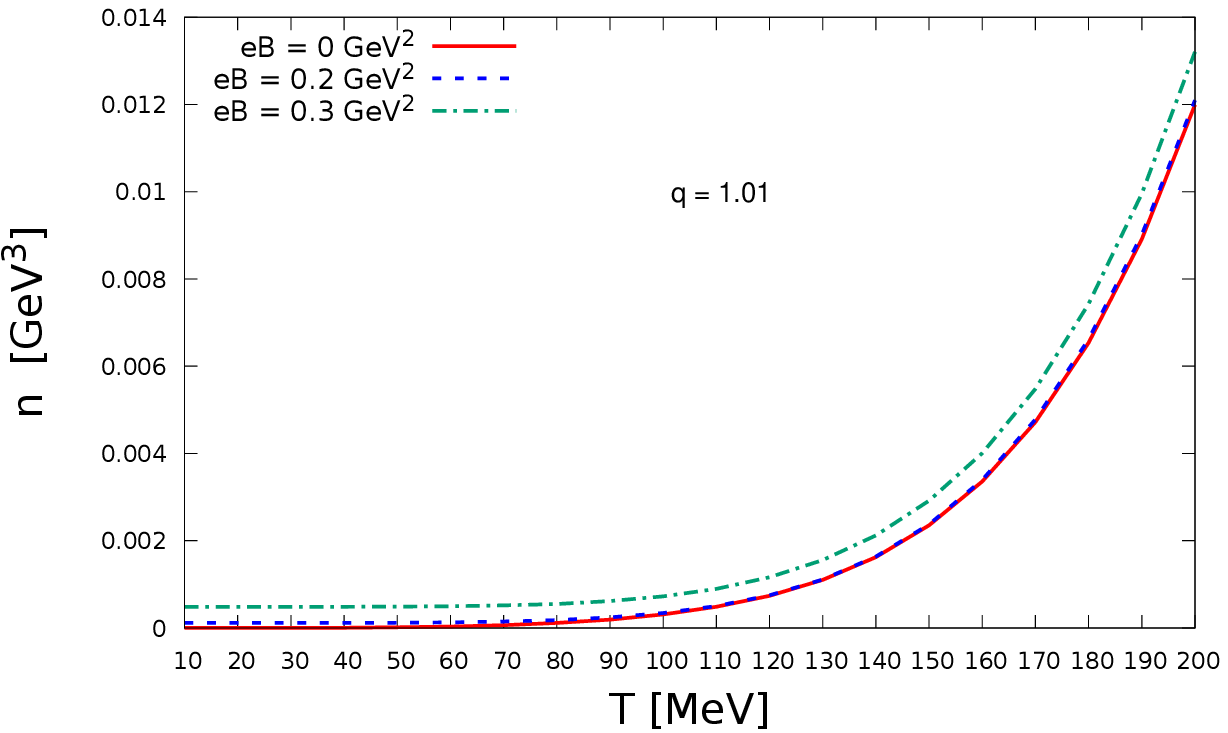}}
  \subfigure[]{\label{fig:dens3}\includegraphics[height=5cm, width=7cm]{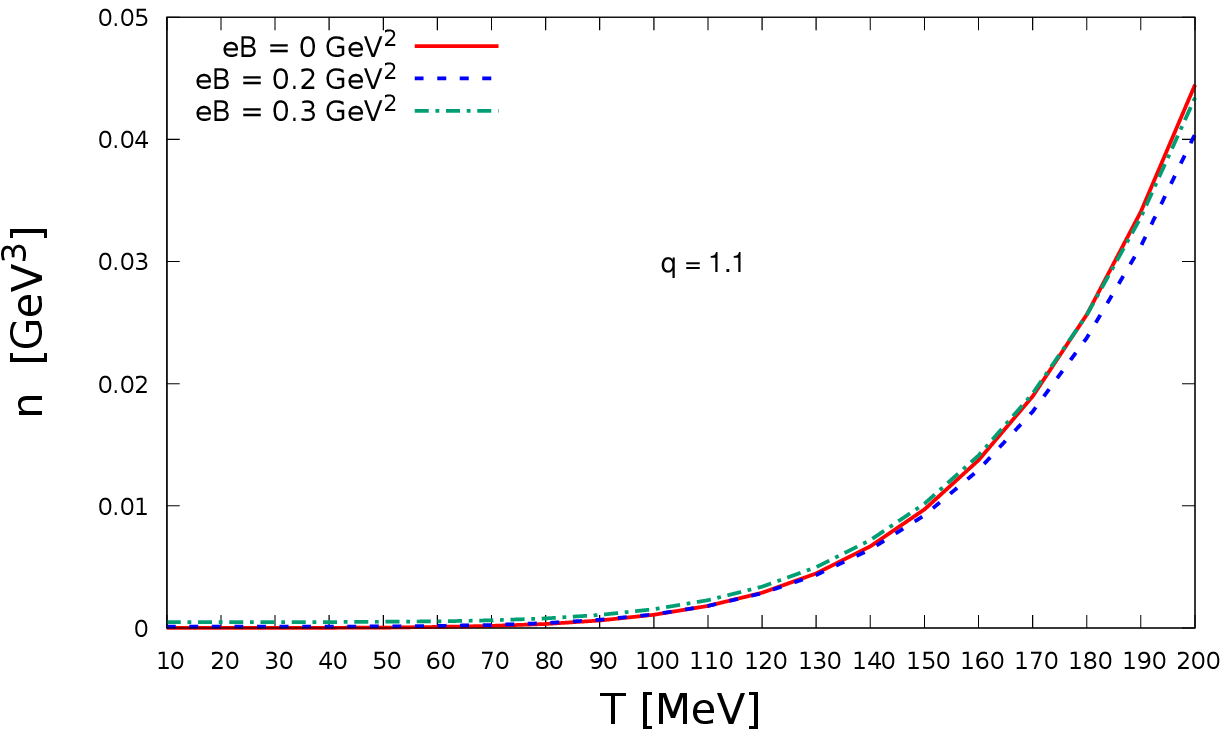}}
  \subfigure[]{\label{fig:dens4}\includegraphics[height=5cm, width=7cm]{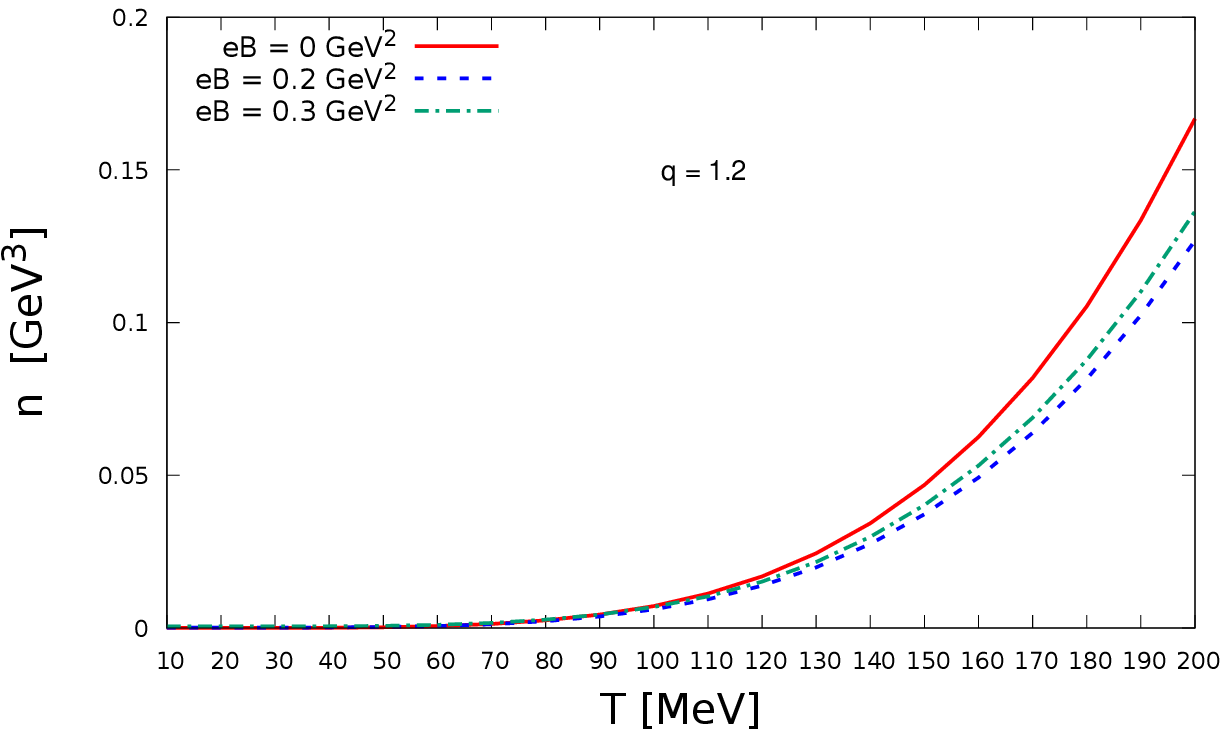}}
  \caption[figtopcap]{The Colored-online number density (left-upper panel) versus the temperature \subref{fig:dens1}, with (red-line),(dashed-blue) and (dotted-dashed green) at ($eB = 0, 0.2, 0.3 \hspace{0.07cm} GeV^{2}$) respectively and at $q = 1.001$. The same conditions are represented in \subref{fig:dens2}, \subref{fig:dens3}, and \subref{fig:dens4} but at $q = 1.01, 1.1, 1.2$ respectively.}
  \label{fig:number_dens}
\end{figure}

  In order to do more investigation, we have studied the number density Fig.\ref{fig:number_dens} with changing the entropic index ($q = 1.001, 1.01, 1.1, 1.2$) with increasing the magnetic field for each case, in which the deviation between zero and non-zero magnetic field occurs at high temperature. his ensures the last result for the presser, energy density and entropy at higher temperature. However we can conclude that increasing the degree of non-extensivity exhibits a reverse magnetic catalysis only at higher temperature 
\section{Conclusion}\label{sec:con}
In the present work, we have explored the effect of the magnetic field on the extensive and non-extensive thermodynamics.
The free energy is separated into two parts, the vacuum and thermal contributions. We have applied the magnetic field on both terms in which the vacuum free energy needs to be regularized according to the reduction in the dimension from 3 to 1. Accordingly, the relativistic energy of the charged particles  has been modified with the Landau levels and polarized spin in z-direction. 
The response of the QCD matter to the magnetic field is examined through the magnetization via the temperature and the magnetic field values. In conclusion the magnetization  showed a negligibly change which slightly increases at low temperature, however it increases with positive values dramatically and rapidly with increasing the magnetic field and the temperature which indicates that, QCD is a paramagnetic matter. All results from both statistics are confronted to the available lattice results at zero and non-zero magnetic field which show a reasonable agreement with the extensive thermodynamics, but overrated in non-extensive thermodynamics at high temperature only, and higher entropic index.

 \appendix
   \section{Appendix A}\label{AppA}
These are the sample of the tables that we have used in our calculations, and includes up to mass $ \sim 10.876 \hspace{0.07cm} GeV$. Table (\ref{tab:hadrons_samp}) contains, the mass code, mass in GeV, width, degeneracy, baryon, strangeness, charmness, bottomness quantum numbers, third component of isospin, charge and the number of decay channels respectively. Table (\ref{tab:hadrons_samp}) also includes information for the daughter in case of the decay of the same particle with "Br" the branching ratio and "Nrs" the number of resonances. In the present work we did not include the decay channels. In addition, the degeneracy appears in table (\ref{tab:hadrons_samp}), then the spin can be extracted from the relation between the degeneracy and spin eq.(\ref{eq:deg}),
\begin{center}
\begin{equation} \label{eq:deg}
g_{i}= 2\hspace{0.02cm}s_{i}+1    
\end{equation}
\end{center}
\begin{table}[ht!]
\centering
\resizebox{\textwidth}{!}{%
\begin{tabular}{|c|l|c|c|c|c|l|c|c|c|c|c|c|l|c|c|}
\hline
MC-Nr (Particle code)&	Name&Mass[GeV]&	Width&dg	&nb&ns&	nc&	nbt&I3&	Q&	dcyChnnl&MC-Nr&daughters&Br$\rightarrow$Daughters-MC-Nrs \\
\hline
\hline
 22	&Gamma	& 0 &	0	& 2&	0	& 0&	0&	0&	1&	0&1&	1&	1&	22\\
211	&	Pion-(+)&		0.13957&	0&	1&	0&	0&	0&	0&	3&	1	&1&1&	1&	211\\
111	&	Pion-(0)&		0.13498	&	0&	1&0&0&	0&0	&3&	0&	1&	1&	1&	111\\
-211&	Pion-(-)&		0.13957	&	0&	1&	0&	0&	0&	0&	3&	-1&	1&	1&	1&	-211\\
\hline
\end{tabular}}

\caption{\label{tab:hadrons_samp} Sample list of hadrons in our PDG \cite{PDG14}.
}
\end{table}

\begin{itemize}
\item Some useful definitions\\
The vacuum part can be solved by using the standard integration in $d$ dimensions (Ref.~\cite{peskin1995introduction})
\begin{eqnarray}\label{eq:intformula}
\int_{-\infty}^{\infty} \frac{d^{d} p}{(2\pi)^{d}} \sqrt{p^{2} + M^{2}} = \frac{1}{(4\pi)^{d/2}} \frac{\Gamma(-1/2-d/2)}{\Gamma(-1/2)} (M^{2})^{1/2+d/2},
\end{eqnarray}
with $\Gamma(-1/2)=-2\sqrt{\pi}$.

\item
The Hurwitz $\zeta$ function is defined as
\begin{eqnarray}
\sum_{k=0}^{\infty} \frac{1}{(v+k)^z} = \zeta(z,v),
\label{eq:Hurwitzdef}
\end{eqnarray}
with the expansion and asymptotic behavior~\cite{dlmf}
\begin{eqnarray}
\zeta(-1+\epsilon/2,v) \approx -\frac{1}{12} -\frac{v^2}{2} + \frac{v}{2} + \frac{\epsilon}{2} \zeta'(-1,v) + \mathcal{O}(\epsilon^2),
\end{eqnarray}
\begin{eqnarray}
\zeta'(-1,v) = \frac{1}{12} -\frac{v^2}{4} + \left( \frac{1}{12} -\frac{v}{2} + \frac{v^2}{2}\right) \log(v) +
\mathcal{O}(v^{-2}).
\label{eq:zeta_asympt}
\end{eqnarray}
\end{itemize}
  \section{Appendix B}\label{AppB}
Beginning with the definition of the momentum again at $x = 0$, $p = \sqrt{\mu^{2}+m^{2}}$
so that the integrand in eq.(\ref{eq:nonext_partF}) will be discontinuous at ($x = 0$). In order to avoid and solve this:-
Firstly, Using the general definition of the average number of particles i as,
\begin{equation}
\langle N \rangle = T \hspace{0.05cm} \frac{\partial}{\partial \mu} log\hspace{0.05cm} \Xi_{q}\Bigg|_{\beta}
\end{equation} 

\begin{equation}
 \langle N \rangle = V \left[ C_{N,q}(\mu,\beta,m) +  \int \frac{d^3p}{(2\pi)^3} \sum_{r=\pm} \Theta(r x) \bigg(\frac{1}{e_q^{(r)}(x) -\xi }\bigg)^{\tilde{q}}  \right]\,, \label{eq:compN}
\end{equation}

where 
\begin{equation}
\tilde{q}=
\begin{cases}
& q \qquad\quad\;\;\, \,,\,\,\,x\geq 0\,, \\
& 2-q \qquad \,,\,\,\,x<0\,.
\end{cases}
\end{equation}
Let the integrand in eq. (\ref{eq:compN}) to be composed of two parts $F^{(-)}, F^{(+)}$, and by converting the three integrals to one integral as;
\begin{eqnarray} 
\int d^{3}p \rightarrow 4 \hspace{0.03cm}\pi\hspace{0.03cm}\int_{0}^{\infty} p^{2}\hspace{0.03cm} dp   \nonumber
\end{eqnarray}
\begin{eqnarray}
\langle N \rangle  &=& T \frac{\partial}{\partial\mu} \left[\int_{0}^{\sqrt{\mu^2-m^2}} dp \, F^{(-)}(p,\mu) + \int_{\sqrt{\mu^2-m^2}}^\infty dp \, F^{(+)}(p,\mu) \right]  \label{eq:generalN}\\
&=&  - \frac{\mu} {\beta \sqrt{\mu^2-m^2}} \left[ F^{(+)}(x=0^+) - F^{(-)}(x=0^-)\right] +  \beta^{-1} \int_0^\infty dp \, \sum_{r=\pm 1} \frac{\partial}{\partial\mu}F^{(r)}(p,\mu) \, \,. \nonumber
\end{eqnarray}
To get the last part in eq. (\ref{eq:generalN}), one can use Leibniz' Rule:

\begin{equation}
\frac{d}{dx} \int_{u(x)} ^{v(x)}  F(x,t)dt = F(x,v(x))\frac{dv}{dx} - F(x,u(x))\frac{du}{dx}+ \int_{u(x)}^{v(x)} \frac{\partial F}{\partial x} dt
\end{equation}
Applying the rule for both parts in the first line of eq. (\ref{eq:generalN}). For the first part, it is straightforward to get,

\begin{equation}
\frac{\partial}{\partial \mu} \int_{0}^{\sqrt{\mu^{2}-m^{2}}}  F^{-}(p,\mu)\hspace{0.03cm} dp = \frac{\mu}{\sqrt{\mu^{2}-m^{2}}}\hspace{0.05cm} F^{-}(\mu,\sqrt{\mu^{2}-m^{2}})+ \int_{0}^{\sqrt{\mu^{2}-m^{2}}} \frac{\partial}{\partial \mu} F^{-}(\mu,p) \hspace{0.03cm}dp
\end{equation}

And the second part in the first line of eq. (\ref{eq:generalN}),
\begin{equation}
\frac{\partial}{\partial \mu} \int_{\sqrt{\mu^{2}-m^{2}}}^{\infty}  F^{+}(p,\mu)\hspace{0.03cm} dp = -\frac{\mu}{\sqrt{\mu^{2}-m^{2}}}\hspace{0.05cm} F^{+}(\mu,\sqrt{\mu^{2}-m^{2}})+ \int_{\sqrt{\mu^{2}-m^{2}}}^{\infty} \frac{\partial}{\partial \mu} F^{+}(\mu,p) \hspace{0.03cm}dp
\end{equation}

Substituting by both parts in eq. eq. (\ref{eq:generalN})
\begin{eqnarray}
\langle N \rangle &=&  \frac{\mu}{\sqrt{\mu^{2}-m^{2}}}\hspace{0.05cm} F^{-}(\mu,\sqrt{\mu^{2}-m^{2}})+ \int_{0}^{\sqrt{\mu^{2}-m^{2}}} \frac{\partial}{\partial \mu} F^{-}(\mu,p) \hspace{0.03cm}dp \label{eq:collectN}\\ &-&\frac{\mu}{\sqrt{\mu^{2}-m^{2}}}\hspace{0.05cm} F^{+}(\mu,\sqrt{\mu^{2}-m^{2}})+ \int_{\sqrt{\mu^{2}-m^{2}}}^{\infty} \frac{\partial}{\partial \mu} F^{+}(\mu,p) \hspace{0.03cm}dp   \, \,. \nonumber
\end{eqnarray}
The discontinuous point at $x = 0$, $p = \sqrt{\mu^{2}+m^{2}}$ , the above eq. (\ref{eq:collectN}) can be rewritten in a compact form,

\begin{equation}
\langle N \rangle =  C_{N,q}(\mu,T,m)+ T  \int _{0}^{\infty} \frac{\partial}{\partial \mu} \sum_{r=\pm} F^{(r)}(\mu,p)dp
\end{equation}
where
\begin{equation}\label{eq:finCterm}
C_{N,q}(\mu,T,m) = -\frac{T\hspace{0.03cm}\mu}{\sqrt{\mu^{2}-m^{2}}}\hspace{0.05cm} \left[  F^{+}(x=0^{+}) - F^{-}(x=0^{-}) \right] 
\end{equation}
Now exploiting eqs. (\ref{eq:exp_nonext}, \ref{eq:log_nonext}, and \ref{eq:nonext_partF}). 
Then one can easily get,
\begin{eqnarray}
F^{+}(x=0^{+}) &=&  \frac{(\mu^{2}-m^{2})}{2\hspace{0.02cm}\pi^{2}} \hspace{0.05cm}\frac{2^{1-q}-1}{1-q} \hspace{0.05cm}\Theta(\mu -m), \nonumber \quad 
 F^{-}(x=0^{-}) = \frac{(\mu^{2}-m^{2})}{2\hspace{0.02cm}\pi^{2}} \hspace{0.05cm}\frac{2^{q-1}-1}{q-1} \hspace{0.05cm}\Theta(\mu -m)  \, \,. \\
 \label{eq:collectN}
\end{eqnarray}
Collect altogether we obtain $\langle N \rangle$.
 
\providecommand{\href}[2]{#2}\begingroup\raggedright

\begin{thebibliography}{100}
\bibitem{letessier2002hadrons} J. Letessier, and J. Rafelski. Hadrons and quark–gluon plasma. Vol. 18. Cambridge University Press, (2002).
\bibitem{Bhalerao14} R. S. Bhalerao, Relativistic heavy-ion collisions, CERN-2014-001 and KEK-Proceedings (2013) 219. ArXiv:1404.3294 [nucl-th].
\bibitem{Csernai} L. P. Csernai, Introduction to relativistic heavy ion collisions,\textsuperscript{\textcopyright} John Wiley and Sons Ltd; Chichester (United Kingdom) (1994).
\bibitem{Chaudhuri} A. K. Chaudhuri, A short course on relativistic heavy ion collisions, \textsuperscript{\textcopyright} IOP Publishing Ltd (2014).
\bibitem{Rafelski} J. Rafelski, Melting Hadrons, Boiling Quarks: From Hagedorn Temperature to Ultra-Relativistic Heavy-Ion Collisions at CERN With a Tribute to Rolf Hagedorn, \textsuperscript{\textcopyright} Springer Cham Heidelberg New York Dordrecht London (2016).‏‏‏
\bibitem{Satz03} V. Magas, H. Satz, Conditions for Confinement and Freeze-Out, Eur.Phys.J. C32 (2003) 115. ArXiv:hep-ph/0308155.
\bibitem{Xu} J. Xu, and C. Ming Ko., Chemical freeze-out in relativistic heavy-ion collisions,  Phys. Lett. B772, (2017) 290.
\bibitem{Cleymans06} J. Cleymans, H. Oeschler, K. Redlich and S. Wheaton, Phys. Rev. C73, 034905 (2006). ArXiv:hep-ph/0511094.
\bibitem{Becattini06} F. Becattini, J. Manninen and M. Gazdzicki, Phys. Rev. C73, (2006) 044905. ArXiv:hep-ph/0511092.
\bibitem{Andronic11} A. Andronic, P. Braun-Munzinger, K. Redlich and J. Stachel, J. Phys. G 38, 124081 (2011). ArXiv:nucl-th/1106.6321.
\bibitem{Altarelli} G. Altarelli, Collider Physics within the Standard Model: A Primer \textsuperscript{\textcopyright} Springer  (2017).
\bibitem{Brambilla} N. Brambilla, et al., QCD and strongly coupled gauge theories: challenges and perspectives, Eur.Phys.J. C74, 10 (2014) 1.
\bibitem{Hagedorn65} R. Hagedorn, Suppl. Nuovo Cimento III, (1965) 147.
\bibitem{Samanta} S. Subhasis, S. Chatterjee, and B. Mohanty, Exploring the hadron resonance gas phase on the QCD phase diagram, J.Phys. G: Nucl.Part.Phys. 46, 6 (2019) 065106.‏
\bibitem{Ruiz Arriola} E. Megias, E. R. Arriola, and L. L. Salcedo, The hadron resonance gas model: thermodynamics of QCD and Polyakov loop, Nucl.Phys.B Proceedings Supplements, 234 (2013) 313.
\bibitem{Hagedorn80} R. Hagedorn, and J. Rafelski, Hot hadronic matter and nuclear collisions, Phys. Lett. B97, 1 (1980) 136.
\bibitem{Shuryak} E. Shuryak, Lectures on nonperturbative QCD (Nonperturbative Topological Phenomena in QCD and Related Theories), arXiv:hep-ph/1812.0150.‏
\bibitem{Sterman} G. Sterman, et al., Handbook of perturbative QCD, Rev. Mod. Phys. 67, 1 (1995) 157.‏‏‏‏‏
\bibitem{Gupta} R. Gupta, Introduction to lattice QCD, Lectures given at the LXVIII Les Houches Summer School "Probing the Standard Model of Particle Interactions", July 28-Sept 5, 1997, arXiv:hep-lat/9807028.
\bibitem{Sourendu} S. Gupta, X. Luo, B. Mohanty, H.G. Ritter, Nu Xu, Scale for the phase diagram of quantum chromodynamics, Science 332, (2011) 1525. arXiv/hep:1105.3934‏
\bibitem{Aoki} Y. Aoki, G. Endrodi, Z. Fodor, S.D. Katz, K.K. Szabo, The order of the quantum chromodynamics transition predicted by the standard model of particle physics, Nature 443, (2006) 675. ArXiv:hep-lat/0611014.
\bibitem{Fodor} G. Endrodi, Z. Fodor, S. D. Katz, and K. K. Szabo, The QCD phase diagram at nonzero quark density
 JHEP, 04 (2011) 001.‏
\bibitem{Kaczmarek} O. Kaczmarek et al.,Phase boundary for the chiral transition in (2+1) -flavor QCD at small values of the chemical potential, Phys. Rev. D83, 014504 (2011).ArXiv:hep-lat/1011.3130.
\bibitem{Bonati} C. Bonati, M. D’Elia, M. Mariti, M. Mesiti, F. Negro, and F. Sanfilippo, Curvature of the chiral pseudocritical line in QCD, Phys. Rev.D90, (2014) 114025.
\bibitem{Bellwied} R. Bellwied, S. Borsányi, Z. Fodor, S. D. Katz, A. Pásztor, C. Ratti, and K. K. Szabó, Fluctuations and correlations in high temperature QCD, Phys.Rev.D92, 11 (2015) 114505.
\bibitem{Bedangadas18} S.Subhasis, and B. Mohanty, Criticality in a hadron resonance gas model with the van der Waals interaction, Phys. Rev. C97, 1 (2018) 015201.‏‏‏‏‏‏‏‏
\bibitem{Vovchenko} V. Volodymyr, M.I. Gorenstein, and H. Stoecker, Van der Waals interactions in hadron resonance gas: from nuclear matter to lattice QCD, Phys. Rev.Lett.118, 18 (2017) 182301.
\bibitem{Agasian2000} N.O. Agasian, Phase structure of the QCD vacuum in a magnetic field at low temperature, Phys. Lett. B488, (2000) 39. 
\bibitem{Endrodi15} G. Endrodi, QCD equation of state at nonzero magnetic fields in the Hadron Resonance Gas model, JHEP  07 (2015) 173. 
\bibitem{Ayalaa16} A. Ayala, C. A.Dominguez, L.A.Hernández, M.Loewe,and R.Zamora, Inverse magnetic catalysis from the properties of the QCD coupling in a magnetic field, Phys.Lett.B759 (2016) 99.
\bibitem{Tawfik16} A. Tawfik, A. Diab, N. Ezzelarab, and Asmaa. G. Shalaby, QCD Thermodynamics and Magnetization in
Nonzero Magnetic Field,  Adv. High. Eng.Phys. 2016, Article ID 1381479, (2016).
\bibitem{Braguta19} V. V. Braguta, M. N. Chernodub, A. Yu. Kotov, A. V. Molochkov and A. A. Nikolaev,  Finite-density QCD transition in a magnetic background field, Phys.Rev. D100, (2019) 114503.
\bibitem{Shuryak84} E. Shuryak, Theory and phenomenology of the QCD vacuum, Phys. Rep.115, (1984) 151.
\bibitem{Bali12} G. S. Bali, F. Bruckmann, G. Endrodi, Z. Fodor, S. D. Katz, S. Krieg, A. Schafer, and K. K. Szabo, The QCD phase diagram for external magnetic fields, JHEP, 2012 (2012) 44. 
\bibitem{Bali14}G. S. Bali, F. Bruckmann, G. Endrodi, S. D. Katz, and A. Schafer, The QCD equation of state in background magnetic fields, JHEP 2014, (2014) 1408.
\bibitem{Buividovich} P. V. Buividovich, M.N. Chernodub, E.V. Luschevskaya, M.I. Polikarpov, Lattice QCD in strong magnetic fields, talk given by M. N. Chernodub at the 10th Workshop on Non-Perturbative QCD, June 8-12, (2009), Paris, France. ArXiv:hep-ph/0909.1808.
\bibitem{Huang} Huang, Xu-Guang, Electromagnetic fields and anomalous transports in heavy-ion collisions: A pedagogical review,  Rep.Prog.Phys. 79 (2016) 076302.
\bibitem{Skokov} Skokov, V. V., A. Yu Illarionov, and V. D. Toneev, Estimate of the magnetic field strength in heavy-ion collisions, Int.J. Mod.Phys.A24, 31 (2009) 5925.
\bibitem{adam_HI} A. Bzdaka, V. Skokov, Event-by-event fluctuations of magnetic and electric fields in heavy ion collisions, Phys.Lett.B710, (2012) 171. ArXiv:hep-ph/1111.1949.
\bibitem{LI_HI} Li Ou and Bao-An Li, Magnetic effects in heavy-ion collisions at intermediate energies, Phys.Rev.C84, (2011) 064605. ArXiv:nucl-th/1107.3192.
\bibitem{Dimitri_HI} D. E.Kharzeev, L.D. McLerran, and Harmen J.Warringa, The effects of topological charge change in heavy ion collisions: “Event by event $\mathcal{P}$ and  $\mathcal{CP}$ violation, Nucl.Phys.A803, (2008) 227. ArXiv:hep-ph/0711.0950.
\bibitem{Mag_Book} D.E. Kharzeev, K. Landsteiner, A. Schmitt, and  Ho-Ung Yee, Strongly interacting matter in magnetic fields, \textsuperscript{\textcopyright} Springer-Verlag Berlin Heidelberg (2013).
\bibitem{Endrodi19} G. Endrődi, M. Giordano, S. D. Katz, T. G. Kovács  F. Pittler, Magnetic catalysis and inverse catalysis for heavy pions, JHEP, 2019 (2019) 7.
\bibitem{Elia18}  M. D’Elia, F. Manigrasso, F. Negro and F. Sanfilippo, QCD phase diagram in a magnetic
background for different values of the pion mass, Phys.Rev.D98 (2018) 054509. ArXiv:hep.lat/1808.07008.
\bibitem{Ilgenfritz13} E.M. Ilgenfritz, M. M$\ddot{u}$ller-Preussker, B. Petersson and A. Schreiber, Magnetic catalysis (and
inverse catalysis) at finite temperature in two-color lattice QCD, Phys.Rev.D89 (2014) 054512. ArXiv:hep-lat/1310.7876.
\bibitem{Bornyakov14} V.G. Bornyakov, P.V. Buividovich, N. Cundy, O.A. Kochetkov and A. Schafer, Deconfinement transition in two-flavor lattice QCD with dynamical overlap fermions in an external magnetic field, Phys.Rev.D90 (2014) 034501. ArXiv:hep-lat/1312.5628.
\bibitem{Bali12R} G. S. Bali, F. Bruckmann, G. Endrődi, Z. Fodor, S. D. Katz, and A. Schäfer, QCD quark condensate in external magnetic fields, Phys.Rev.D86, 071502(R).
\bibitem{Voronyuk} V. Voronyuk, V. D. Toneev, W. Cassing, E. L. Bratkovskaya, V. P. Konchakovski, and S. A. Voloshin, Electromagnetic field evolution in relativistic heavy-ion collisions
, Phys. Rev. C83, (2011) 054911. ArXiv:nucl.th/1103.4239.
\bibitem{Becattini04} F. Becattini, M. Gaydzicki, A. Keranen, J. Manninen, R. Stock, Chemical equilibrium study in nucleus-nucleus collisions at relativistic energies, Phys. Rev. C69 (2004) 024905.
\bibitem{Abhijit2020} G. Kadam, S. Pal, A. Bhattacharyya, Interacting hadron resonance gas model in magnetic field and the fluctuations of conserved charges, J.Phys.G47 (2020) 125106. ArXiv:hep-ph/1908.10618v2.
\bibitem{Shalaby16} Asmaa G. Shalaby, Extensive and non-extensive thermodyamics, Acta Phyica.Polon B47 (2016).
\bibitem{Rozynek19} Jacek Rożynek and Grzegorz Wilk, Nonextensive Quasiparticle Description of QCD Matter, Symmetry, 11(3), (2019) 401.
\bibitem{Parta12} N.K. Patra, T. Malik, D. Sen, T.K. Jha, and H. Mishra, An Equation of State for Magnetized Neutron Star Matter and Tidal Deformation in Neutron Star Mergers, Astrophys.J. 900 (2020) 1, 49.
\bibitem{Shalaby2021} Asmaa G. Shalaby, Vasilis K. Oikonomou, and Gamal G. L. Nashed, Non-Extensive Thermodynamics Effects in the Cosmology of f(T) Gravity., Symmetry, 13(1), 75 (2021).
\bibitem{boltzmann} L. Boltzmann, Kinetische Theorie II, (1970) 115.
\bibitem{gibbs} J. W. Gibbs, Elementary principles in statistical mechanics: developed with especial reference to the rational foundations of thermodynamics, \textsuperscript{\textcopyright} C. Scribner's sons (1902).
\bibitem{reif2009fundamentals} F. Reif, Fundamentals of statistical and thermal physics, \textsuperscript{\textcopyright} Waveland Press (2009).
\bibitem{Bohr69} A. Bohr and B. R. Mottelson, Nuclear Structure \textsuperscript{\textcopyright} W. A Benjamin, Inc. (1969).
\bibitem{Gupta81} B.K. Jennings, L. Satpathy, and S. Das Gupta, Approximate method of including finite particle number effects in relativistic heavy ion collisions, Phys.Rev.C24, (1981) 440.
\bibitem{Chase95} K.C. Chase and A.Z. Mekjian, Exact methods for expectation values in canonical fragmentation models
, Phys.Rev.C52, (1995) 2339.
\bibitem{Gupta98} S. Das Gupta and A.Z. Mekjian, Phase transition in a statistical model for nuclear multifragmentation, Phys. Rev.C57, (1998) 1361.
\bibitem{Keranen} A. Keranen, F. Becattini, Chemical factors in canonical statistical models for relativistic heavy ion collisions, Phys.Rev.C65, (2002) 044901, Erratum Phys.Rev.C68 (2003) 059901.
\bibitem{Abhijit} A. Bhattacharyya, S. K. Ghosh, R.Ray, and S. Samanta. Exploring effects of magnetic field on the Hadron Resonance Gas, EPL 115 (2016) 62003. Arxiv:hep-ph/1504.04533.
\bibitem{Tawfik_Shalaby} A. Tawfik, Loutfy I. Abou-Salem, Asmaa G. Shalaby, M. Hanafy, A. Sorin, O. Rogachevsky, and W. Scheinast, Particle production and chemical freezeout from the hybrid UrQMD approach at NICA energies, Eur. Phys. J. A52 (2016) 324.
\bibitem{endrodi} G. Endrödi, QCD equation of state at nonzero magnetic fields in the Hadron Resonance Gas model
, JHEP 04, 023 (2013).
\bibitem{Andersen14} Jens O. Andersen, William R. Naylor, Anders Tranberg, Phase diagram of QCD in a magnetic field: A review, Rev. Mod. Phys. 88, 025001 (2016). ArXiv:hep-ph/1411.7176v3
\bibitem{Kapusta06} J. Kapusta and C. Gale, Finite-Temperature Field Theory: Principles and Applications. Cambridge monographs on mechanics and applied mathematics, \textsuperscript{\textcopyright} Cambridge University Press, (2006).
\bibitem{Fraga08} E. S. Fraga and A. J. Mizher, Chiral transition in a strong magnetic background, Phys.Rev. D78 (2008) 025016. ArXiv:hep-ph/0804.1452.
\bibitem{Landau77} L. Landau and E. Lifshits, Quantum Mechanics: Non-Relativistic Theory. Course of theoretical physics (Vol 3.) (Landau, L. D, 1908-1968)  \textsuperscript{\textcopyright} Pergamon Press, (1977).
\bibitem{PDG14} K. Olive et al., Summary tables of particle properties, Chin. Phys. C38, 090001 (2014).
\bibitem{George75} G. Leibbrandt, Introduction to the technique of dimensional regularization
, Rev. Mod. Phys. 47 (1975) 849.
\bibitem{Menezes08} D.P.Menezes, M. Benghi Pinto, S.S. Avancini, A. Perez Martinez, and C. Providencia, Quark matter under strong magnetic fields in the Nambu--Jona-Lasinio Model, Phys.Rev.C79 (2009) 035807. ArXiv:nucl-th/0811.3361.
\bibitem{GellMann95} M. Gell-Mann, The Quark and the Jaguar: Adventures in the Simple and the Complex, Macmillan (1995).
\bibitem{Holovatch17} Y. Holovatch, R. Kenna, and S. Thurner, Complex systems: physics beyond physics
, Eur.J.Phys.38 (2017) 023002. ArXiv:Physics/1610.01002.
\bibitem{Tsallis09} C. Tsallis, Introduction to nonextensive statistical mechanics: approaching a complex world  \textsuperscript{\textcopyright} Springer Science  BusinessMedia (2009).
\bibitem{Tsallis11} C. Tsallis, Nonextensive statistical mechanics: Applications to high energy physics, Eur. Phys.J. Web of Conferences, 13 (2011) 05001.
\bibitem{Plastino93} A. Plastino and A. Plastino, Stellar polytropes and Tsallis' entropy, Phys.Lett.A174 (1993) 384.
\bibitem{Plastino99} A. Plastino and A. R. Plastino, Tsallis Entropy and Jaynes' information theory formalism, Braz. J.Phys. 29  (1999) 50.
\bibitem{Megias14} E. Megias, D. P. Menezes, and A. Deppman, Nonextensive thermodynamics with finite chemical potentials and protoneutron stars, Eur. Phys.J. Web of Conferences, 80 (2014) 00040. ArXiv:hep-ph/1407.8044.
\bibitem{Abe2001} S. Abe and Y. Okamoto, Nonextensive statistical mechanics and its applications, Vol. 560 \textsuperscript{\textcopyright} Springer Science  Business Media, (2001).
\bibitem {Tsallis88} Tsallis, C. Possible generalization of Boltzmann-Gibbs statistics. J. Stat.Phys. 52 (1988) 479.
\bibitem{Umarov} S. Umarov, C. Tsallis, S. Steinberg, On a q-Central Limit Theorem Consistent with Nonextensive Statistical Mechanics, Milan J. Math. 76, 307 (2008).
\bibitem{Rath19} R. Rath, S. Tripathy, B. Chatterjee, R. Sahoo, S. Kumar Tiwari, and A. Nath, Violation of Wiedemann-Franz Law for Hot Hadronic Matter created at NICA, FAIR and RHIC Energies using Non-extensive Statistics, Eur. Phys. J. A 55, 125 (2019). ArXiv:hep-ph/1902.07922.
\bibitem{Tsallis02} C. Tsallis, Nonextensive statistical mechanics: A brief review of its present status, Annals of the Brazilian Academy of Sciences, 74(3) (2002) 393. ArXiv:cond-mat/0205571.
\bibitem{TsallisMath} C. Tsallis, Computational applications of nonextensive statistical mechanics, J.Comp.Appl.Math. 227 (2009)  51.
\bibitem {Megias15} E. Megias, D.P. Menezes, A. Deppman, Non extensive thermodynamics for hadronic matter with finite chemical potentials, Physica A421 (2015) 15.
\bibitem{peskin1995introduction} M. Peskin and D. Schroeder, An Introduction To Quantum Field Theory \textsuperscript{\textcopyright} Advanced Book Program. Basic Books, (1995).
\bibitem{dlmf} http://dlmf.nist.gov.
 \end{thebibliography}
\end{document}